\documentclass[preprint,10pt,twocolumn, twoside]{revtex4}%
\usepackage{amsmath}%
\usepackage{amsfonts}%
\usepackage{amssymb}%
\usepackage{graphicx}
\usepackage[caption=false]{subfig}
\usepackage{basic_math}
\usepackage{qm}
\usepackage{hyperref}
\hypersetup{pdfpagemode=UseNone} 
\usepackage[normalem]{ulem}

\def\colorfig{}%
\begin{document}
\title{
Quantum enhancement of sensitivity achieved by photon-number-resolving detection in the dark port of a two-path interferometer operating at high intensities
}
\author{Jun-Yi Wu}
\email{junyiwuphysics@gmail.com}
\author{Norifumi Toda}
\author{Holger F. Hofmann}
\affiliation{Graduate School of Advanced Sciences of Matter, Hiroshima University, Kagamiyama 1-3-1, Higashi Hiroshima 739-8530, Japan}
\begin{abstract}
  It is shown that the maximal phase sensitivity of a two-path interferometer with high-intensity coherent light and squeezed vacuum in the input ports can be achieved by photon-number-resolving detection of only a small number of photons in a dark output port. It is then possible to achieve the quantum Cram\'{e}r-Rao bound of the two-path interferometer using only the field displacement dependence of the photon number statistics in the single mode output of the dark port represented by a field-displaced squeezed vacuum state. We find that, at small field displacements, it is not sufficient to use the average photon number as the estimator, indicating that an optimal phase estimation depends critically on measurements of the precise photon number. We therefore analyze the effect of detection efficiency on the Fisher information and show that there is a transition from low robustness against photon losses associated with quantum interference effects at low field displacements to high robustness against photon losses at high field displacements. The transition between the two regimes occurs at field shifts proportional to the third power of the squeezing factor, indicating that squeezing greatly enhances the phase interval in which quantum effects are relevant in optimal phase estimations using photon resolving detectors. The case under study could thus be understood as a `missing link' between genuine multiphoton interference and the straightforward suppression of noise usually associated with squeezed light.
\end{abstract}
\maketitle

\section{Introduction}
\label{sec::introduction}
  Two-path interferometers are highly sensitive to small phase shifts as demonstrated e.g. by their application to gravitational wave detection in the Laser Interferometer Gravitational-Wave Observatory (LIGO) \cite{LIGO2011-GW, LIGO2013-GW, LIGO2016-GW}.
  A quantum enhancement of this high phase sensitivity can be achieved by feeding a coherent light field into one of the input ports and a squeezed vacuum state into the other\cite{Caves1981-QNoiseSqSt, LangCaves2013-OptQEnhance, YurkeMcCallKlauder1986-SU2SU11Interf, HilleryMlodinow1993-InterfNSqSt, Paris1995-CohSqNSqSqInterf, BarnettMaitre2003, AtamanPredaIonicioiu2018-PhSnsng1MdN2Md}.
  In this case, the phase shift is proportional to the change in the photon number difference in the outputs, and the squeezing reduces the photon number fluctuations of this difference, increasing the signal to noise ratio for the phase estimate.
  It therefore seems to be unnecessary to apply the complete theoretical analysis that is usually needed to achieve the quantum Cram\'{e}r-Rao (QCR) bound of a non-classical input state\cite{BraunsteinCaves1994-StDistGeoQSt}.
  However, as we will show in the following, the average photon number difference between the two output ports is not the optimal estimator in the situation where almost all of the photons exit the interferometer in only one of the two output ports, leaving the other port dark.
  This situation is similar to the optimal phase estimation procedure for combinations of squeezed vacuum with comparably weak coherent light, which can result in photon statistics dominated by genuine multiphoton coherences \cite{PezzeSmerzi2008-MZIHsnbgLmtCSV, Hofmann2009-PureStHsnbgLmt, OnoHofmann2010-PhEst}.
  In the present case, the few photons left in the dark port exhibit the highly nonclassical photon statistics of a field-displaced squeezed vacuum state, requiring an optimal estimator that is based on precise photon counting in the dark port output only.
  It has already been pointed out previously that the optimal estimator for a combination of coherent state and squeezed vacuum in the output ports is the parity of odd and even photon numbers\cite{SeshadreesanEtAlDowling2011-ParityDtctnHsnbgLmt}.
  In the case of sufficiently high intensity in the coherent field, we can describe the effect of finite phase shifts by a field-displacement of the squeezed vacuum and characterize the gradual change of the optimal estimation strategy from parity to average photon number.

  It is interesting to observe that the full sensitivity of the high photon number in the bright output port can be obtained from a precise measurement of the few photons in the dark port, without detecting any of the photons in the bright output port.
  However, the need to resolve the precise photon number of the dark port requires special photon-number-resolving detectors (PNRDs)\cite{MillerEtAlSergienko2003-PNRD, KhouryEtBouwmeester2006-PNRDOfNnlnrInterf, LitaMillerNam2008-PNRD, WildfeuerEtAlDowling2009-PNRDFPInterf, MatekoleEtAlDowling2017-RmTmpPhNrDtctn} operating at high detection efficiency.
  Any photon losses in the PNRDs will result in a significant reduction of the sensitivity from the QCR bound.
  In the regime where the average photon number is the optimal estimator, photon losses will have less effect on the sensitivity.
  The fragility of the quantum sensitivity enhancement in the presence of photon losses is therefore a characteristic quantum property that distinguishes the suppression of noise by squeezing from the multiphoton interferences that can only be accessed by the precise photon statistics obtained from PNRDs.
  It may thus be of interest to take a closer look at the photon counting statistics of the dark port output and its relation with the robustness of sensitivity against photon losses.

  The rest of the paper is organized as follows.
  In Section \ref{sec::ph_stat_para_est}, we introduce the description of the phase dependent photon number statistics in the dark port regime and compare the performance of the optimal phase estimators with the average photon number as a function of the phase bias of the interferometer.
  In Section \ref{sec::ph_stat_est_sq_st}, we analyze the precise photon statistics of the displaced squeezed states and identify the characteristic quantum interference effects on their photon number statistics.
  In Section \ref{sec::trans_ph_stat_sq_st}, we identify a critical quadrature displacement value and show that the quantum interference effects become minor effects for displacement greater than this critical value.
  In Section \ref{sec::reduction_FI_in_lossy_PNRD}, photon losses in the PNRDs are introduced, and the reduction of phase sensitivity caused by these losses is analyzed. It is shown that the robustness against photon losses increases as the optimal estimator changes from a complicated photon number dependence at low bias phases to the average photon number at higher bias phases.
  Section \ref{sec::conclusion} concludes the paper.

\section{
Phase estimation in the dark port regime
}
\label{sec::ph_stat_para_est}

  \begin{figure}[t]
    \includegraphics[width=0.4\textwidth]{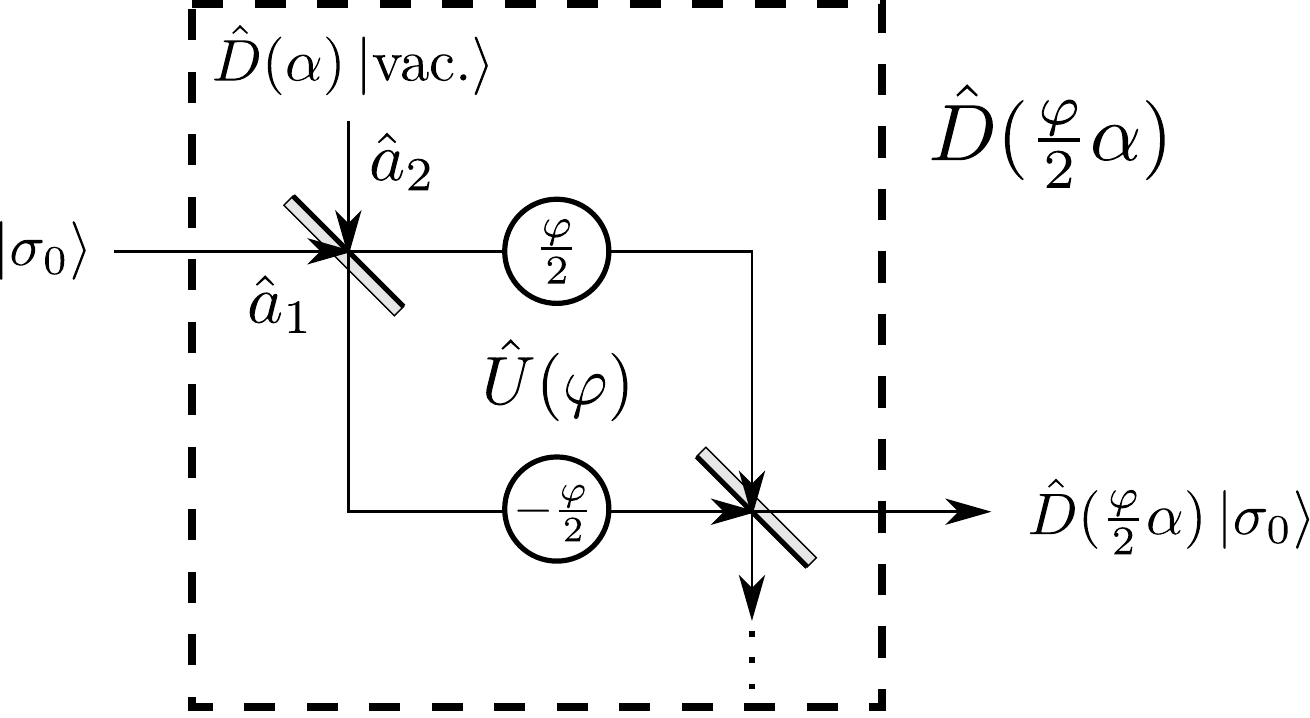}
    \caption{
    Illustration of the field displacement of a low-intensity input state $\ket{\sigma_{0}}$ in a two-path interferometer using a high-intensity coherent state $\ket{\alpha}$.
    For a small phase shift $\varphi\ll1$, a small part $\alpha \varphi/2$ of the coherent amplitude is transferred to the output mode of the low-intensity input state, resulting in a unitary displacement operation acting on the input state $\ket{\sigma_{0}}$.
    }
    \label{fig::scheme_phase_est_dark_port}
  \end{figure}

  Figure \ref{fig::scheme_phase_est_dark_port} shows a schematic representation of a two-path Mach-Zehnder interferometer commonly used in optical quantum metrology.
  The phase difference between the two internal paths is $\varphi$.
  The input modes $\hat{a}_{1,2}$ are transformed into the output modes by a unitary operator $\hat{U}(\varphi)$ that transforms the modes according to
  \begin{equation}
  \label{eq::mod_transf}
    \hat{U}\colvec{2}{\hat{a}_{1}}{\hat{a}_{2}}\hat{U}^{\dagger}
    =
    \left(
      \begin{array}{cc}
        \cos(\frac{\varphi}{2}) & -\sin(\frac{\varphi}{2}) \\
        \sin(\frac{\varphi}{2}) & \cos(\frac{\varphi}{2}) \\
      \end{array}
    \right)
    \colvec{2}{\hat{a}_{1}}{\hat{a}_{2}}.
  \end{equation}
  A squeezed vacuum state $\ket{\sigma_{0}}$ is input in mode $1$, and a high-intensity coherent state $\ket{\alpha}$ is input in mode $2$.
  In the following, we will define the single mode phases so that $\alpha$ is a real number and is phase locked to the squeezed quadrature of $\ket{\sigma_{0}}$.
  It is then possible to express the two-mode input state using a two-mode displacement operator $\hat{D}(0, \alpha)$ acting on the product of vacuum and squeezed vacuum,
  \begin{equation}
    \ket{\sigma_{0}}\otimes\ket{\alpha}
    =
    \hat{D}(0, \alpha) \ket{\sigma_{0};\text{vac.}}.
  \end{equation}
  The interference of the input states can now be expressed by separate transformations of the displacement and the partially squeezed two-mode vacuum,
  \begin{equation}
    \hat{U}(\varphi)\ket{\sigma_{0};\alpha}
    =
    \hat{U}(\varphi)\hat{D}(0, \alpha)\hat{U}^{\dagger}(\varphi) \;\hat{U}(\varphi)\ket{\sigma_{0};\text{vac.}}.
  \end{equation}
  The transformation of the displacement operator changes the amplitude of the displacement according to the transformation of the modes given by Eq. \eqref{eq::mod_transf},
  \begin{equation}
    \hat{U}(\varphi)\hat{D}(0,\alpha)\hat{U}^{\dagger}(\varphi)
    =
    \hat{D}\left(\sin(\frac{\varphi}{2})\alpha, \cos(\frac{\varphi}{2})\alpha\right).
  \end{equation}
  For small phases $\varphi\ll1$, the field displacement of mode $1$ is approximately $(\alpha \varphi/2)$, while the displacement of mode $2$ is nearly unchanged.
  The transformation of the partially squeezed vacuum $\ket{\sigma_{0};\text{vac.}}$ describes photon transfers from the squeezed vacuum $\ket{\sigma_{0}}$ to the true vacuum $\ket{\text{vac.}}$.
  However, the low photon number of the squeezed vacuum makes such a transfer highly unlikely at small phases $\varphi\ll1$, so the unitary transformation leaves the partially squeezed vacuum approximately unchanged.
  \begin{equation}
    \hat{U}(\varphi)\ket{\sigma_{0};\text{vac.}}
    \approx
    \ket{\sigma_{0};\text{vac.}}
  \end{equation}
  Since we can neglect the entanglement generated by the transformation at small phases, the output is a product state of a squeezed vacuum state displaced by a field quadrature of $x=\alpha\varphi/2$ and a coherent state of amplitude $\alpha$.
  The coherent state in the bright output port is nearly unchanged by the small phase $\varphi$ and therefore carries no phase information whatsoever.
  It is therefore possible to completely discard the intense light in the bright output port and obtain an optimal phase estimate from photon detection in the dark output port only.

  The output state in the dark port is given by a quadrature-displaced squeezed vacuum,
  \begin{equation}
  \label{eq::sq_st}
    \ket{\sigma(x)} = \hat{D}(x)\ket{\sigma_{0}} ,
  \end{equation}
  where the field displacement is proportional to the phase $\varphi$ with a proportionality factor given by half the amplitude of the high-intensity coherent field,
  \begin{equation}
  \label{eq::prop_phi_alpha}
    x = \frac{\alpha}{2} \varphi.
  \end{equation}
  It is therefore possible to obtain an estimate of a small shift of the phase $\varphi$ by estimating the value of a small variation of the field displacement $x$ of the quadrature-displaced squeezed vacuum.
  The proportionality in Eq. \eqref{eq::prop_phi_alpha} determines the relation between the phase sensitivity $1/\Delta\varphi^{2}$ and the quadrature sensitivity $1/\Delta x^{2}$,
  \begin{equation}
  \label{eq::x_phi_sens_n}
    \frac{1}{\Delta \varphi^{2}} = \frac{\alpha^{2}}{4}\frac{1}{\Delta x^{2}},
  \end{equation}
  where $\Delta \varphi^{2}$ and $\Delta x^{2}$ describe the variances of the estimates obtained from the measurement results.
  In the following, we will thus focus on the estimation of small variations in the field displacement $x$ parameterizing the single mode output state in the dark port.

  The information on small variations in the quadrature displacement $x$ is contained in the conditional photon number statistics $p_{n}(x)=|\braket{n|\sigma(x)}|^{2}$.
  As shown in \cite{Hofmann2009-PureStHsnbgLmt}, it is possible to achieve the QCR bound with these probabilities by choosing an optimized estimator.
  The QCR bound is given by the quantum Fisher information $\mathcal{H}_{F}(x)$ of the displaced squeezed vacuum state $\ket{\sigma(x)}$,
  \begin{equation}
  \label{eq::FI_and_QFI_pure_st}
    1/\Delta x^{2} \le \mathcal{H}_{F}(x)
  \end{equation}
  For a parameterized unitary transformation, the quantum Fisher information is given by the uncertainty of the generator of the unitary.
  In the case of a displacement of the quadrature $\hat{X}$, this is the uncertainty of the conjugated quadrature $\hat{Y}$,
  \begin{align}
  \label{eq::QFI_pure_st}
    \mathcal{H}_{F} = 16\Delta Y^{2},
  \end{align}
  where $\Delta Y^{2} = \braket{\hat{Y}^{2}}-\braket{\hat{Y}}^{2}$.
  Since this bound can be achieved using photon number measurements\cite{Hofmann2009-PureStHsnbgLmt}, the QCR bound is given by the classical Fisher information of the probability distribution $p_{n}(x)$,
  \begin{equation}
    \sum_{n}\frac{(\partial_{x}p_{n})^{2}}{p_{n}} =16\Delta Y^{2}.
  \end{equation}
  This means that, even though the probability distribution $p_{n}(x)$ depends on the displacement $x$, the optimal phase sensitivity always achieves the same value given by the displacement independent QCR bound of the squeezed vacuum.

  \begin{figure}[t]
    \includegraphics[width=0.45\textwidth]{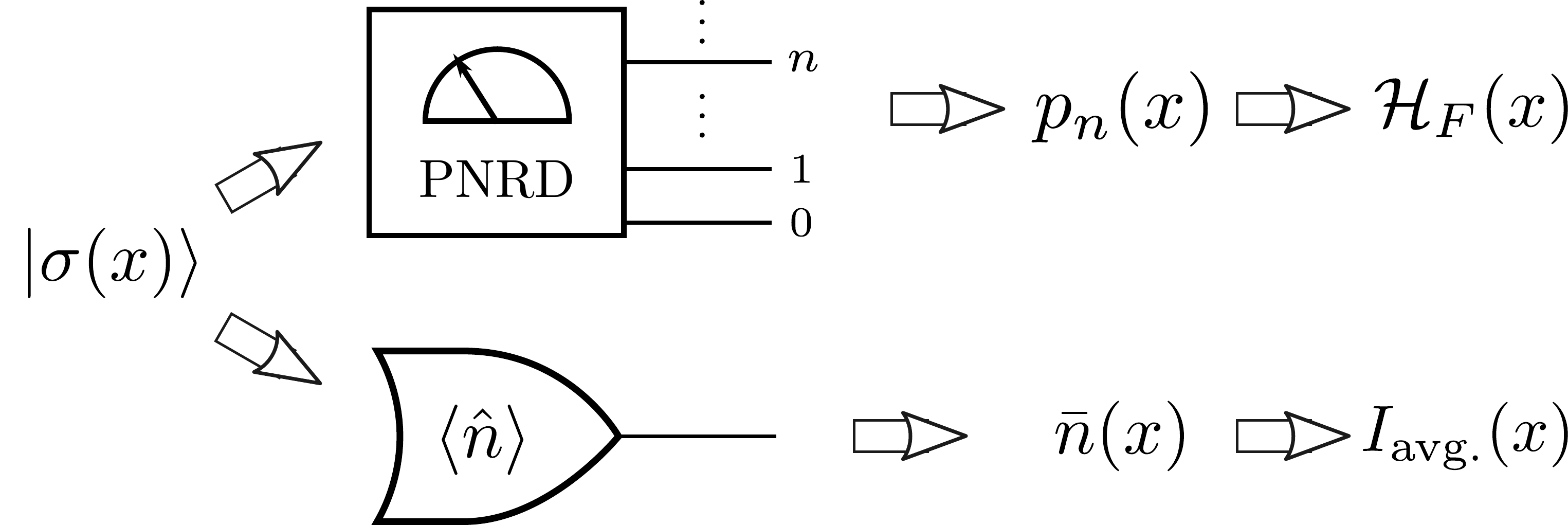}
    \caption{
    Comparison of estimation strategies for small variations in the $x$-quadrature displacement.
    The sensitivity is equal to the Fisher information of the quantum state $\mathcal{H}_{F}$ for a PNRD measurement resolving the complete distribution of photon numbers $n$.
    Alternatively, it is possible to average the photon numbers over a large number of measurements, as shown in the lower part of the figure.
    From the measured average photon number $\bar{n}(x)$, one can estimate the value of $x$ with a resolution equal to $I_{\text{avg.}}$, which is generally lower than the quantum Fisher information $\mathcal{H}_{F}$ as shown in Eq. \eqref{eq::I_cl}.
    }
    \label{fig::scheme_PNRD_2}
  \end{figure}
  On the practical side, the problem is that the optimal phase estimator depends on the details of $p_{n}(x)$ and may be very sensitive to changes in the precise distribution caused by photon losses or similar detection errors.
  To get an idea of how robust the sensitivity is against photon losses, it may be helpful to compare the QCR bound with the sensitivity achieved by using the average photon number as an estimator, as shown in Fig. \ref{fig::scheme_PNRD_2}.
  In this case, the sensitivity is given by the ratio of the squared $x$-derivative of the average photon number $\bar{n}$ and the photon number uncertainty $\Delta n^{2} = \braket{\hat{n}^{2}}-\braket{\hat{n}}^{2}$,
  \begin{equation}
  \label{eq::I_cl_derivation}
    I_{\text{avg.}}(x)
    = \frac{(\partial_{x}\bar{n})^{2}}{\Delta n^{2}}.
  \end{equation}
  For the displaced squeezed vacuum state $\ket{\sigma(x)}$, the sensitivity is given by the quantum Fisher information multiplied with a sigmoid function,
  \begin{gather}
    I_{\text{avg.}}(x)
    =
    \mathcal{H}_{F}\frac{x^{2}}{x^{2}+\chi_c^{2}}
    \label{eq::I_cl}
  \end{gather}
  where the critical displacement $\chi_{c}$ defines the width of the sigmoid function.
  At the displacement $x=\chi_{c}$, the sensitivity $I_{\text{avg.}}$ achieves half of the quantum Fisher information $\mathcal{H}_{F}$.
  For displacements much larger than $\chi_{c}$ the sensitivity approaches the QCR bound indicating that the average photon number is the optimal estimator.
  It should be noted that this estimate corresponds to homodyne detection of the squeezed vacuum, where the large amplitude displacement acts as a local oscillator field, converting the field quadrature fluctuation into photon number fluctuations in the outputs. The critical displacement $\chi_{c}$ therefore marks the transition from a photon number sensitive detection to a detection of interference effects of the quadrature component in phase with the displacement field.
  The critical displacement $\chi_{c}$ can be determined from the quantum statistics of the squeezed vacuum.
  Specifically, $\chi_{c}$ depends on the photon number uncertainty $\Delta n_{0}$ and the $\hat{X}$-quadrature uncertainty $\Delta X_{0}$ of the undisplaced squeezed vacuum $\ket{\sigma_{0}}$,
  \begin{equation}
  \label{eq::chi_def}
    \chi_c = \frac{\Delta n_{0}}{2\Delta X_{0}}.
  \end{equation}
  For an squeezed vacuum input $\ket{\sigma_{0}}$ with a squeezing factor $r$, the critical displacement $\chi_{c}$ is given by
  \begin{equation}
  \label{eq::chi_value}
    \chi_{c} = \frac{1}{2\sqrt{2}}(e^{3r}-e^{-r}),
  \end{equation}
  which is a third-order function of the squeezing factor $e^{r}$.
  The critical displacement corresponds to a critical phase $\varphi_{c}$ in the two-path interferometer, which depends on the amplitude $\alpha$ of the coherent light in the bright input port:
  \begin{equation}
    \varphi_{c} = \frac{1}{\sqrt{2}}\frac{e^{3r}-e^{-r}}{\alpha}.
  \end{equation}
  This relation shows that the amount of squeezing together with the intensity of the coherent input determine the range of bias phases $[-\varphi_{c}, \varphi_{c}]$ for which the average photon number is not a very good estimator.
  Since the coherent input is much stronger than the squeezed vacuum input, the critical phase is typically very small.
  However, the increase with the third power of the squeezing factor indicates that the range of bias phases $[-\varphi_{c},\varphi_{c}]$ can be significantly enlarged by stronger squeezing at a fixed coherent amplitude.

  The reason why the sensitivity of the average photon number estimate is lower than the optimal estimate is the non-classical statistics of photon numbers $n$ in the dark port.
  Therefore $\varphi_{c}$ provides a condition for the range of bias phases with highly nonclassical photon number statistics in the dark output port.
  We will hence take a closer look in the next section at the precise distributions of photon number as a function of the $x$ displacement resulting from bias phases $\varphi$ .

\section{
Nonclassical photon number statistics in the dark port of the interferometer
}
\label{sec::ph_stat_est_sq_st}

  \begin{figure*}[th]
    \centering
    \includegraphics[width=0.9\textwidth]{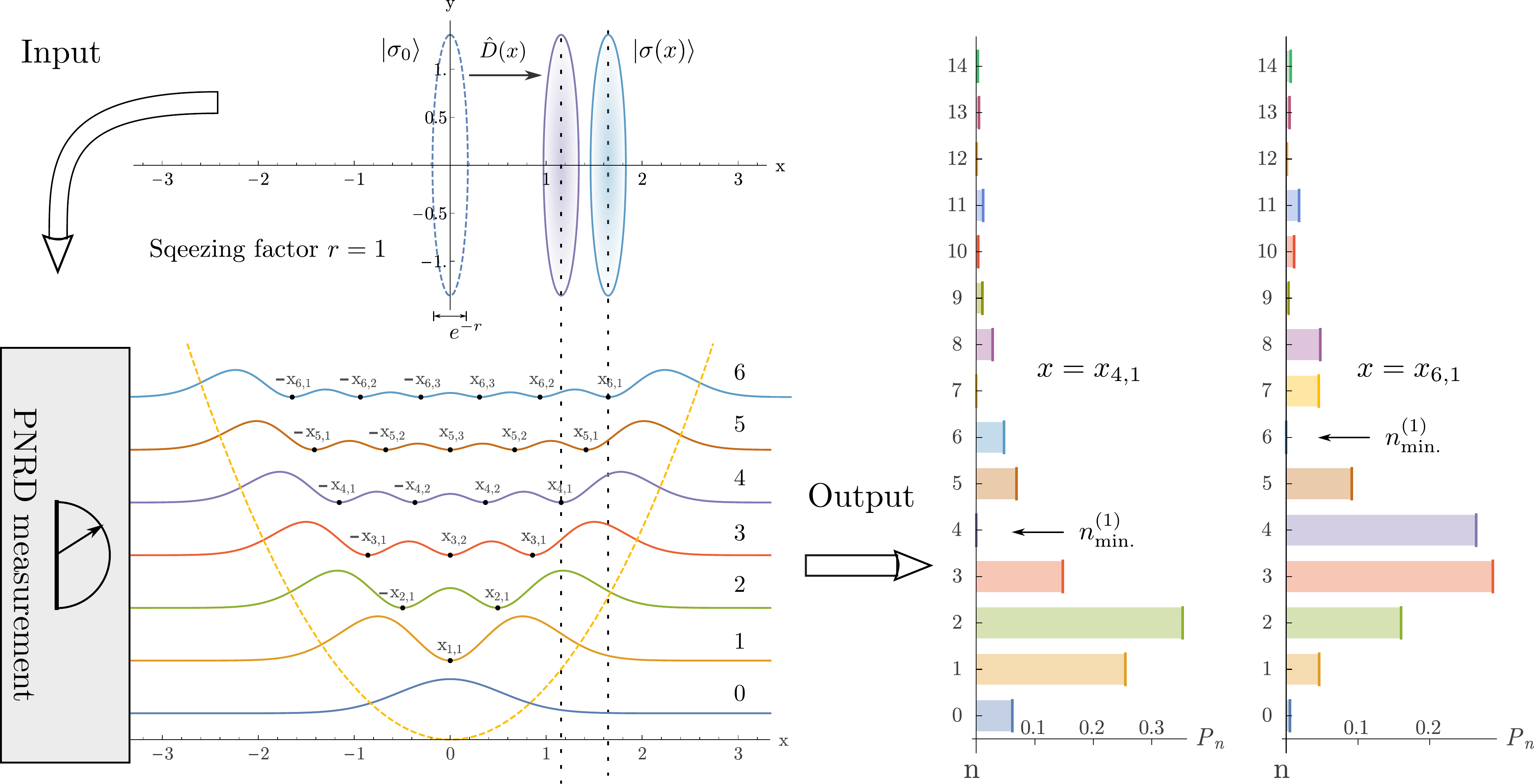}
    \caption{\colorfig
      Illustration of the displacement dependence of the photon statistics observed using PNRDs in the dark port of the two-path interferometer for a squeezing parameter of $r=1$.
      The displacement of the phase space distribution is shown on the upper left-hand side of the figure.
      The displacement dependence of the detection probabilities is shown on the lower left.
      An offset proportional to $(n+1/2)$ is used to distinguish the different photon numbers.
      The parabola shows the value of $x^{2}$ in units of the offset, indicating that probability rapidly drops to zero for displacement larger than $\sqrt{n+1/2}$.
      The zero points of the distributions are marked with $x_{n,k}$ as explained in the text.
      The right hand side of the figure shows the photon number distributions for $x=x_{4,1}\approx1.16$ and $x=x_{6,1}\approx1.65$, where the lowest minima of the photon number distribution are found at $n=4$ and $n=6$, respectively.
    }\label{fig::PNRD_of_sq_st}
  \end{figure*}

  In this section, we characterize the photon statistics of the $x$-displaced squeezed state $\ket{\sigma(x)}$ and show that the oscillations observed in the photon number distributions can be explained as an effect of quantum interference between two different phases of the single mode oscillation.

  The photon statistics $p_{n}(x) = |\braket{n|\sigma(x)}|^{2}$ of a displaced squeezed state can be obtained from the inner products of photon number states $\ket{n}$ and the displaced squeezed state $\ket{\sigma(x)}$,
  \begin{equation}
  \label{eq::inner_prod_sq_st}
    \braket{n|\sigma(x)}
    =
    \left(\sqrt{1-\gamma^{2}}\frac{\gamma^{n}}{n!}\right)^{\frac{1}{2}}
    H_{n}(2\zeta x)e^{-\frac{2\gamma}{1+\gamma}(\zeta x)^{2}},
  \end{equation}
  where $H_{n}$ is the probabilists' Hermite polynomial, and $\gamma$ and $\zeta$ are functions of the squeezing parameter $r$ given by
  \begin{align}
    \gamma = \tanh(r)
    &&\text{ and }&&
    \zeta = \frac{1}{\sqrt{1-e^{-4r}}}
  \end{align}
  Figure \ref{fig::PNRD_of_sq_st} illustrates the photon statistics of a displaced squeezed state with a squeezing parameter of $r=1$.
  On the lower left, the dependence of detection probabilities on displacements is shown for photon numbers from $0$ to $6$.
  The Gauss-Hermite functions given by Eq. \eqref{eq::inner_prod_sq_st} describes oscillation of the probability amplitudes that result in displacement values $x_{n,k}$ where the probability $p_{n}$ drops to zero.
  As shown in the figure, we define $x_{n,k}$ as the
  displacement of the $k$-th zero point for $p_{n}(x)$ counted from large displacements to the center.
  The zero point displacements $x_{n,k}$ can be related to the zero point arguments $z_{n,k}$ of the probabilists' Hermite polynomials $H_{n}(z)$ by
  \begin{equation}
  \label{eq::zero_sq_st_exact}
    x_{n,k} = \frac{1}{2}\sqrt{1-e^{-4r}}z_{n,k}.
  \end{equation}
  This relation illustrates the rapid transition from $r=0$ where the only zero points are found at $x=0$ to a situation where the zero points are close to $z_{n,k}/2$ observed for $\hat{X}$-quadrature eigenstates.

  \begin{figure}
    \centering
    \includegraphics[width=0.49\textwidth]{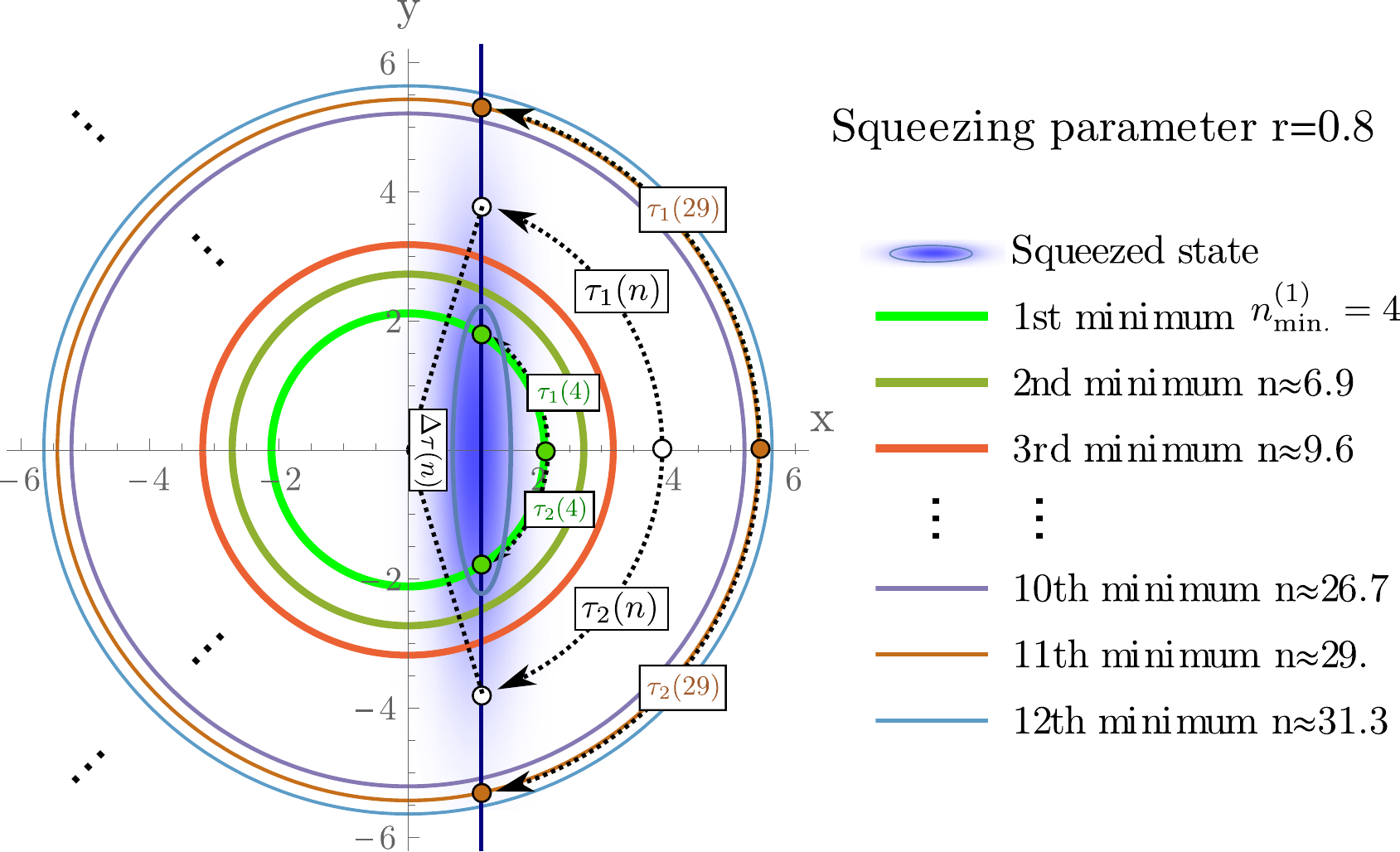}
    \caption{\colorfig Explanation of the origin of multiphoton interference fringes in the photon number distribution $p_{n}(x)$ for a quadrature displacement of $x = x_{4,1} \approx 1.14$ and a squeezing parameter of $r=0.8$.
    The quantum state $\ket{\sigma(x)}$ is indicated by a straight line parallel to the $y$-axis at the displacement $x$.
    The actual phase space extent of the Gaussian Wigner function of $\ket{\sigma(x)}$ is illustrated by the blue-shaded region.
    The dotted arcs indicate single mode phase shifts, with $\Delta\tau(n)$ describing the phase difference between the two phases $\tau_{1}(n)$ and $\tau_{2}(n)$ that intersect the central quadrature value $x$ along a circle of photon number $n$.
    The solid circles show the positions of minima caused by destructive interferences in the photon number distribution $p_{n}$.
    Note that these minima are not necessarily at integer photon numbers.
    As shown in Eq. \eqref{eq::n_sep_n_phase}, the product of the arc $\Delta\tau(n)$ and the photon number difference $\Delta\nu$ between two consecutive circles is approximately equal to $2\pi$.
    Since $\Delta\tau(n)$ gradually increases from an initial value of about $2\pi/3$ to a final value of $\pi$, the separation $\Delta\nu$ between consecutive minima decreases from about $3$ to an asymptotic limit of $2$ as photon number increases.
    }
    \label{fig::q_cl_regimes}
  \end{figure}

  When considering the displacement dependence of the whole photon number distribution, the locations of the minima $n_{\text{min.}}$ in the probability distribution $p_{n}$ appear to move to higher photon numbers as the displacement increases.
  This is illustrated on the right-hand side of Fig. \ref{fig::PNRD_of_sq_st}, where the zero point at $n=4$ observed at a displacement of $x_{4,1}$ seems to move to $n=6$ as the displacement increases to $x_{6,1}$.
  In general, at a displacement of the value $x=x_{n,1}$, the corresponding photon number distribution $p_{n}$ has the first probability minimum $n_{\text{min.}}^{(1)} = n$ following the first peak.
  By considering the $n$-dependence at a fixed displacement of $x$, the multiphoton interference fringes given by $p_{n}(x)$ appear as interference fringes in photon number $n$ following the first minimum.
  These interference fringes are a series of oscillations with decreasing periodicity approaching a period of $2$ photons for high $n$, which corresponds to the periodicity observed for quantum states with a specific photon number parity such as the squeezed vacuum without displacement or a NOON state superposition.

  It is possible to explain this interference pattern in the photon number distributions as a superposition of two quantum state components separated by a unitary single mode phase shift of $\Delta\tau(n)$ in phase space\cite{HibinoEtAlHofmann2018-StatFromAction}.
  Figure \ref{fig::q_cl_regimes} shows a phase space illustration of the photon number dependence of this single mode phase shift.
  For sufficiently high squeezing parameters, the displaced squeezed state can be characterized by the displacement $x$, which is related to two phase values $\tau_{1,2}(n)$ satisfying $x=\sqrt{n+1/2}\cos\tau_{1,2}(n)$ for each photon number $n$.
  As explained in \cite{HibinoEtAlHofmann2018-StatFromAction}, the separation $\Delta\nu$ between two consecutive minima in $p_{n}$ can be derived from the phase difference $\Delta\tau(n) = \tau_{1}(n) - \tau_{2}(n)$ using the following relation,
  \begin{equation}
  \label{eq::n_sep_n_phase}
    \Delta\nu \approx \frac{2\pi}{\Delta\tau(n)}.
  \end{equation}
  As shown in Fig. \ref{fig::q_cl_regimes}, the phase difference $\Delta\tau(n)$ has its smallest value at the minimum with lowest photon number $n_{\text{min.}}^{(1)}$ and gradually increases with increasing photon number.
  According to Eq. \eqref{eq::n_sep_n_phase}, this means that the separation $\Delta\nu$ has the largest value between the first and second minima and decrease as photon numbers increase.
  For an extremely large photon number, the phase difference $\Delta\tau$ is approximately $\pi$, so that the separations between minima turn into a constant periodicity of $2$ photons.
  The photon number distribution $p_{n}$ therefore exhibits a series of quantum interference fringes with a clear periodic pattern explained by the phase differences $\Delta\tau(n)$ between the intersections of the quantum state $\ket{\sigma(x)}$ with states of photon number $n$ in phase space.

  Of particular interest is the location of the first minimum $n_{\text{min.}}^{(1)}$.
  This minimum marks the high photon number end of the first and highest peak in the photon number distribution $p_{n}$.
  Interference effects only become relevant when this first minimum is located well within the overall photon number distribution.
  If the separation between the first minimum and the average photon number is much larger than the photon number uncertainty, the photon number distribution will appear as a single peak that can be approximated by a Gaussian distribution.
  It is therefore possible to distinguish two separate regimes based on a comparison between the separation of the first minimum and the average photon number and the photon number uncertainty of the quantum state.
  In the following, we will show that this transition is identical to the transition at which the average photon number becomes a good estimator for the phase estimate.

\section{Transition from quantum interference to single peak statistics}
\label{sec::trans_ph_stat_sq_st}

  The exact photon number distribution $p_{n}$ in the dark port for a given displacement $x$ is given by the absolute squares of the probability amplitudes $\braket{n|\sigma(x)}$ shown in Eq. \eqref{eq::inner_prod_sq_st}.
  This inner product can be expressed by appropriately modified Gauss-Hermite functions.
  It is well-known that these Gauss-Hermite functions can be approximated by a product of an envelope function and a modulation with a phase of S, e.g., by using the WKBWentzel-Kramers-Brillouin approximation to find the solution of the harmonic oscillator eigenstates.
  Using this approximation, we can write the photon number distribution for $n+1/2>x^{2}$ as follows:
  \begin{equation}
  \label{eq::q_inter_approx}
    p_{n}(x) \approx 2\rho(n,\zeta x)\cos^{2}(S(n,\zeta x)-\frac{\pi}{4}),
  \end{equation}
  where the coarse grained probability distribution is given by
  \begin{equation}
  \label{eq::energy_envelop}
    \rho(n,x) = \frac{1}{\sqrt{2\pi(n+\frac{1}{2}-x^{2})\Delta Y^{2}}}e^{-\frac{n+\frac{1}{2}-x^{2}}{2\Delta Y^{2}}}
  \end{equation}
  and the quantum phase $S(n,\zeta x)$ in the interference term is given by the intersecting area between the circle of photon number $n$ and the straight line representing an $X$-quadrature eigenstate $\ket{\zeta x}$ as illustrated in Fig. \ref{fig::q_cl_regimes}.
  If the optical phase $\tau$ is not too large, this area can be approximated by
  \begin{equation}
    S(n(y),\zeta x) = S(y,\zeta x) \approx \frac{2y^{3}}{3 \zeta x},
  \end{equation}
  where $y$ is the $Y$-quadrature of the phase space point defined by a photon number $n$ and an $X$-quadrature of $\zeta x$,
  \begin{equation}
  \label{eq::y_n_x}
    y = \sqrt{n+\frac{1}{2} - (\zeta x)^{2}}.
  \end{equation}
  $S(y,\zeta x)$ describes the quantum phase responsible for the interference pattern in $p_{n}(x)$ as shown in Eq. \eqref{eq::q_inter_approx}.
  The first destructive quantum interference occurs at $S(y,\zeta x) = 3\pi/4$, with further minima occurring after each increase by $\pi$.
  It is therefore possible to evaluate the number $k$ of destructive quantum interferences that occur up to a photon number corresponding to a $Y$-quadrature value of $y$ by
  \begin{equation}
  \label{eq::S_x_y}
    k = \left\lfloor \frac{1}{\pi}S(y,\zeta x) + \frac{1}{4} \right\rfloor.
  \end{equation}
  As $x$ increases, the number $k$ of destructive quantum interferences below the value of $y$ decreases, until $\zeta x=(8y^{3}/9\pi)$, where the first destructive interference is located at $y$.
  In general, the $Y$-quadrature value of the $k$-th minimum is determined by
  \begin{equation}
    y_{\text{min.}}^{(k)}
    =
    \left((4k-1)\frac{3\pi}{8}\zeta x\right)^{\frac{1}{3}}.
  \end{equation}
  This relation indicates that the locations of the minima of the photon number distribution $p_{n}$ are monotonically increasing with the field displacement $x$.

  The statistics of the coarse grained distribution $\rho(n,x)$ is easier to understand if it is transformed into a distribution of $y$ values given by Eq. \eqref{eq::y_n_x},
  \begin{equation}
    \rho(n,x)\frac{dn}{dy} = \frac{2}{\sqrt{2\pi \Delta Y^{2}}}e^{-\frac{y^{2}}{2\Delta Y^{2}}},
  \end{equation}
  where the factor of $2$ indicates that there are two $y$-values for each value of $n$.
  We can therefore use the Gaussian statistics of $y$ to estimate the probability of finding photon numbers below or above the first occurrence of destructive quantum interference.
  This estimate can tell us how important quantum interferences are in the photon number distribution.

  For $y = \Delta Y$, the approximate probability of finding photon numbers with higher $y$ values is $32\%$.
  If $y=\Delta Y$ is the $Y$-quadrature value of the first minimum, $68\%$ of the detected photon numbers will be lower than the first minimum and will therefore be found within a single peak in the probability distribution $p_{n}$.
  This condition is satisfied when the displacement is $x=\chi_{_{\Delta Y}}$ with
  \begin{equation}
  \label{eq::x_deltaY}
    \chi_{_{\Delta Y}}=\frac{8\Delta Y^{3}}{9\pi\zeta} .
  \end{equation}
  Since the critical displacement $\chi_{c}$ given in Eq. \eqref{eq::chi_value} is also proportional to the $\Delta Y^{3}$, the displacement $\chi_{_{\Delta Y}}$ can be directly related to this critical displacement.
  The relation obtained by comparing Eq. \eqref{eq::x_deltaY} and \eqref{eq::chi_value} is
  \begin{equation}
    \chi_{_{\Delta Y}} = \frac{2\sqrt{2}}{9\pi} \zeta\chi_{c}.
  \end{equation}
  For $\zeta \approx1$, the ratio is approximately given by $\chi_{c}\approx 10 \chi_{_{\Delta Y}}$.
  This means that the transition of the sensitivity for average photon number estimate is closely related to the increase in probability of photon numbers belonging to the first peak of the photon number distribution.

  \begin{figure}
    \centering
    \includegraphics[width=0.5\textwidth]{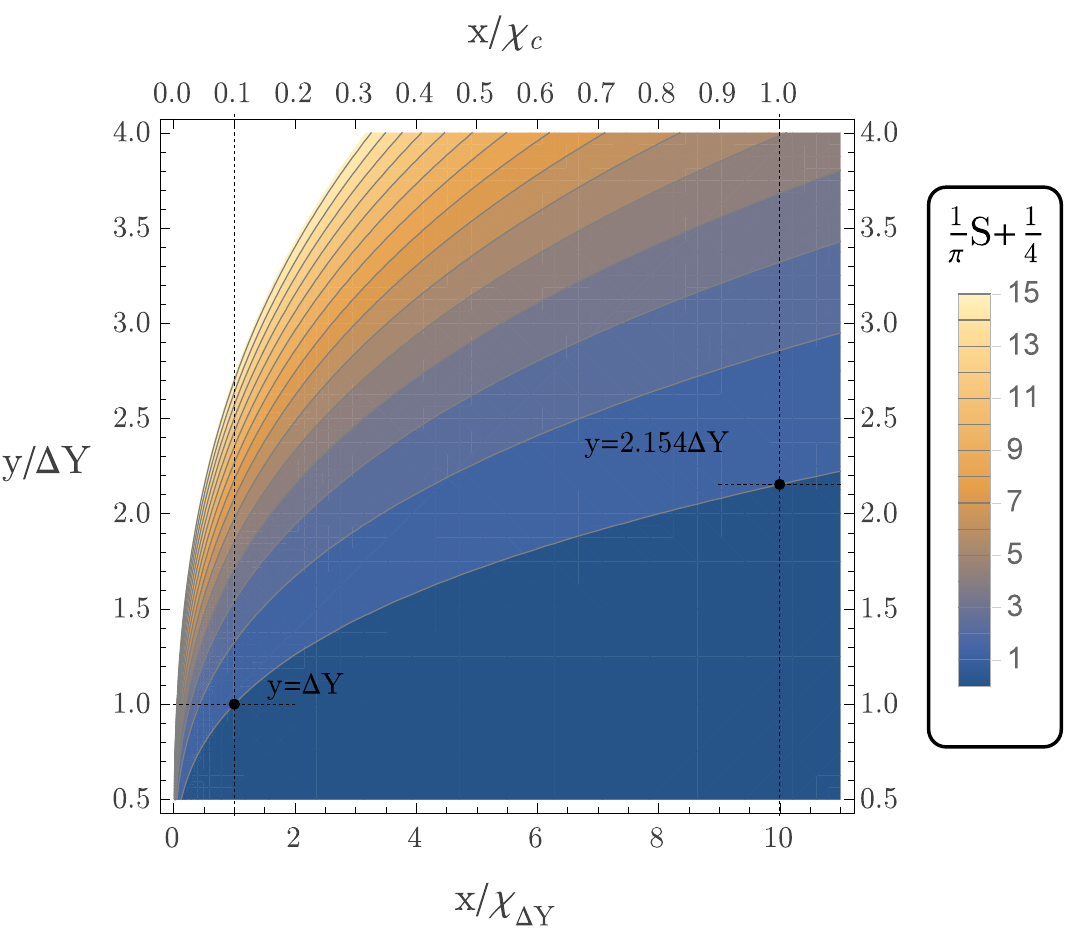}
    \caption{The number of destructive quantum interferences at $Y$-quadratures smaller than a value of $y$ for an $x$-displaced squeezed state $\ket{\sigma(x)}$ with a large squeezing factor, as indicated by a contour plot of $S/\pi +1/4$.
    The lower scale gives the displacement $x$ in units of $\chi_{_{\Delta Y}}$ marking the occurrence of the first minimum at $y = \Delta Y$, and the upper scale gives $x$ in units of the critical displacement $\chi_{c}$ describing the sensitivity of average photon estimates.
    }\label{fig::first_min_DY_x}
  \end{figure}

  Figure \ref{fig::first_min_DY_x} shows the positions of minima in the photon number distribution in the form of a contour plot of $(S/\pi+1/4)$ as given by Eq. \eqref{eq::S_x_y}.
  At $x=\chi_{_{\Delta Y}}$, the first minimum of the photon number distribution is found at $y=\Delta Y$, corresponding to a probability of approximately $68\%$ of photon number outputs being found in the first peak.
  As displacements $x$ increase, the position of the first minimum then moves to higher values of $y$, so that the total probability of photon number outputs being found in the first peak increases.
  This tendency is the reason for the increase of sensitivity for average photon number estimate given by Eq. \eqref{eq::I_cl}.
  At the critical displacement $x=\chi_{c}$, the value of $y$ for the first minimum is
  \begin{equation}
    y_{\text{min.}}^{(1)} = \sqrt[3]{\frac{9\pi}{2\sqrt{2}\zeta}}\Delta Y \approx 2.154 \Delta Y.
  \end{equation}
  This means that about $95\%$ of the photon number distribution is in the first peak.

  At displacements larger than $\chi_{c}$, most of the photon number distribution is well below the first minimum, and the shape of the distribution is approximately a Gaussian.
  It is therefore possible to evaluate the distance between average photon number and the first minimum in terms of the photon number uncertainty $\Delta n$ to get a clear idea of how well the photon number distribution approximates a Gaussian.
  For $x=\chi_{c}$, the photon fluctuation is given by
  \begin{equation}
    \Delta n  = 2\Delta Y^{2},
  \end{equation}
  while the average photon number is
  \begin{equation}
    \bar{n} = \chi_{c}^{2} + \Delta Y^{2} - \frac{1}{2}.
  \end{equation}
  The first minimum is determined to be
  \begin{equation}
    n_{\text{min.}}^{(1)}
    =
    (\zeta\chi_{c})^{2}+ \left(y_{\text{min.}}^{(1)}\right)^{2} - \frac{1}{2}
    \approx
    \bar{n} + 2.07\Delta n.
  \end{equation}
  This means that for all displacements $x\ge\chi_{c}$, the first minimum is separated from the average photon number by a distance of more than $2\Delta n$.
  In this regime, quantum interferences only modulate the probability $p_{n}$ of photon numbers $n$ that are larger than $\bar{n}+2 \Delta n$, which is only a minor modification of the low-probability higher-photon-number tail of the Gaussian distribution.
  It is then possible to approximate the total photon number distribution with a single peak statistics given by the Gaussian distribution
  \begin{equation}
  \label{eq::cl_approx}
    p_{n}(x) \approx
    \frac{1}{\sqrt{2\pi}\Delta n}e^{-\frac{(n-\bar{n})^{2}}{2\Delta n^{2}}},
  \end{equation}%
  which explains why the average photon number estimate achieves the QCR bound.

  \begin{figure*}[t]
    \centering
    \includegraphics[width=0.99\textwidth]{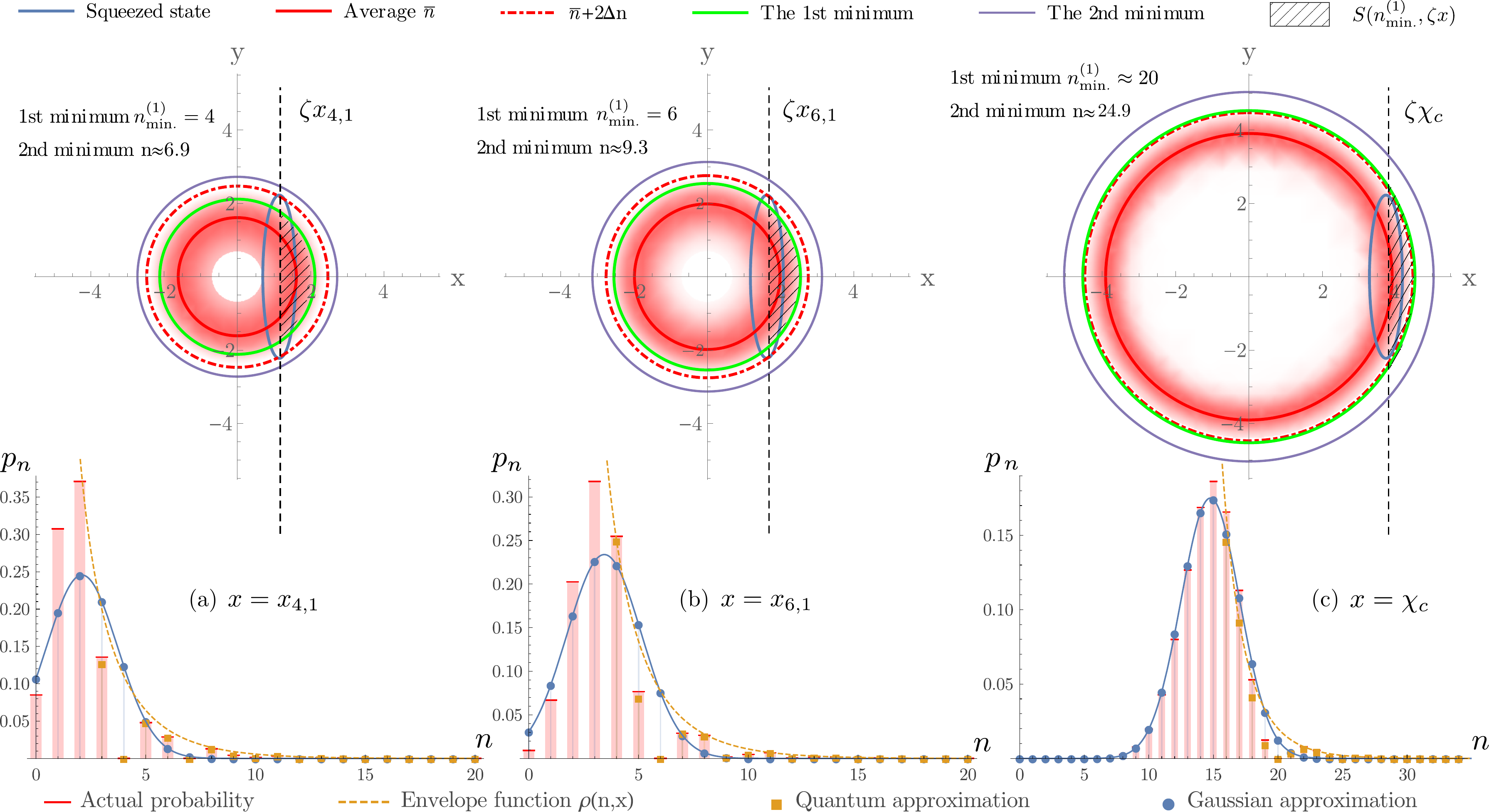}
    \caption{\colorfig
      Explanation of quantum interference fringes in photon number distributions of displaced squeezed states with a squeezing parameter of $r=0.8$.
      In the upper part, the first minimum $n_{\text{min.}}^{(1)}$ (green circle) and the regime $[0,\bar{n}+2\Delta n]$ (red shading inside the red dash-dotted line) for different quadrature displacements are shown in phase space.
      The dashed line represents the $\zeta x$-quadrature.
      In the lower part, the actual probability distributions $p_{n}$ of displaced squeezed states obtained from Eq. \eqref{eq::inner_prod_sq_st} are plotted as histograms.
      The approximate quantum interference fringes given by Eq. \eqref{eq::q_inter_approx} are plotted with orange squares, which are obtained from the orange dashed envelop function $2\rho(n,\zeta x)$ modulated by a squared cosine function with a phase of $(S(n,\zeta x)-\pi/4)$.
      The approximate Gaussian distribution for $\bar{n}$ and $\Delta n$ given by Eq. \eqref{eq::cl_approx} are plotted using blue circles.
      Panel (a) shows the photon statistics for a small displacement of $x=x_{4,1}\approx1.14$. The photon number distribution is well approximated by Eq. \eqref{eq::q_inter_approx}, but is quite different from the Gaussian distribution.
      Panel (b) shows the photon statistics for an intermediate displacement of $x=x_{6,1}\approx1.62$. The probabilities for photon numbers greater than $4$ are well approximated by Eq. \eqref{eq::q_inter_approx}.
      The probabilities of photon number from $0$ to $3$ are roughly given by the Gaussian, with a maximal deviation at $3$ photons.
      Panel (c) shows the photon statistics for a large displacement of $x=\chi_{c}\approx3.74$.
      In this limit, the approximation of Eq. \eqref{eq::q_inter_approx} converges on the Gaussian approximation of Eq. \eqref{eq::cl_approx}.
      Only a slight deviation from the Gaussian exists around $n=20$.
    }\label{fig::prob_approx}
  \end{figure*}

  \bigskip

  Figure \ref{fig::prob_approx} illustrates the photon number statistics of displaced squeezed states for a squeezing parameter of $r=0.8$ at different displacements.
  In the upper part of the figure, we show the relation between the first minimum $n_{\text{min.}}^{(1)}$ (green circle) and the photon numbers $[0,\bar{n}+2\Delta n]$ (red shading inside the red dash-dotted line) in phase space.
  The corresponding probability distribution $p_{n}$ is shown in the lower part of the figure.
  One can observe that for a small displacement value $x=x_{4,1}\approx1.14$, the quantum interferences have significant effects on the photon number statistics.
  As the displacement increases, the quantum interference fringes shift to higher photon numbers.
  This shift is faster than the combined increase of average photon number and photon number uncertainty given by $\bar{n}+2\Delta n$.
  For a quadrature at the critical displacement $\chi_{c}$, the whole photon number statistics can be well approximated by the single peak Gaussian function given in Eq. \eqref{eq::cl_approx} with only a small modification by the interference fringes around the photon number $n=20$.

  \begin{figure}[t]
    \centering
    \includegraphics[width=0.5\textwidth]{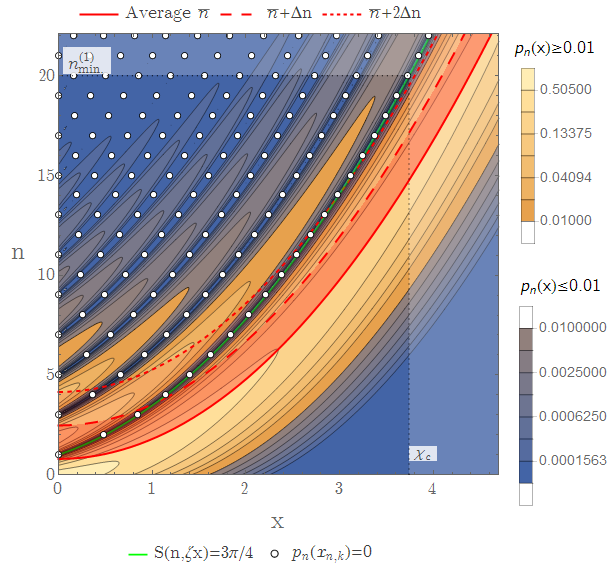}
    \caption{\colorfig
      Transition from quantum interference to Gaussian distribution.
      The contour plot shows the photon number distribution $p_{n}(x)$ of squeezed states with $r=0.8$, where the photon number dependence is mathematically interpolated between discrete photon numbers to give a more intuitive image of the photon number dependence.
      The red solid line indicates the average photon number $\bar{n}$.
      The magnitude of photon number uncertainty $\Delta n$ is illustrated by the dashed and dotted lines, showing $\bar{n}+\Delta n$ and $\bar{n}+2\Delta n$, respectively.
      The region between $\bar{n}$ and $\bar{n}+2\Delta n$ is highlighted in red.
      The white circles mark the zero points $x_{n,k}$ of $p_{n}(x)$.
      The green line marks the interpolation of the lowest photon number zero points associated with $x_{n,1}$, indicating the photon number at which quantum interference starts to occur.
      As the displacement $x$ increases, the zero points shift out of the regime between $\bar{n}$ and $\bar{n}+2\Delta n$ and the photon statistics approaches a Gaussian as given by Eq. \eqref{eq::cl_approx}.
    }\label{fig::ph_stat_chi}
  \end{figure}

  Figure \ref{fig::ph_stat_chi} illustrates the transition between quantum interference and Gaussian statistics by a contour plot of $p_{n}(x)$ for a squeezing parameter of $r=0.8$.
  The contour plot makes it easy to see the shift of quantum interference fringes to higher photon numbers as the displacement $x$ increases.
  The average photon number and the photon number uncertainty are indicated by the red lines.
  It is easy to see that the interference fringes shift to higher photon numbers faster than the sums of average photon number and photon number uncertainty given by $\bar{n}+\Delta n$ and $\bar{n}+2\Delta n$.
  For $x\ge\chi_c$, the quantum interference fringes are all found above photon numbers of $\bar{n}+2 \Delta n$.

  The results of this section confirm that the reason for the transition in the sensitivity of the average photon number estimate discussed in Sec. \ref{sec::ph_stat_para_est} is the transition from quantum interference in the photon number distribution to a single-peaked Gaussian distribution.
  The average photon number becomes an inefficient estimator for $x<\chi_{c}$ as quantum interference effects modify the $x$-gradients of the probability distribution $p_{n}(x)$.
  Efficient estimators then require precise detection of photon numbers,
  since the value of the estimator may be quite different for neighboring photon numbers.
  It is therefore necessary to use PNRDs to achieve the QCR bound, and the actual phase resolution that can be achieved will be very sensitive to photon losses in the detection process.
  We will therefore quantify the effects of photon losses that limit the detection efficiencies of PNRDs on the Fisher information that can be  extracted by photon detection in the dark output port of the two-path interferometer.

\section{
Reduction of Fisher information by photon losses in PNRDs
}

  In a realistic PNRD measurement, photon losses in the detection process modify the photon statistics and reduce the visibility of quantum interferences in the photon number distribution.
  The Fisher information contributed by quantum interferences is therefore sensitive to photon losses, especially at the displacements $x=x_{n,k}$, where the photon number $n$ has zero probability in a perfect PNRD measurement.
  A similar phenomenon has been observed in the experiments of phase sensing using NOON states and Holland-Burnett states \cite{OkamotoEtAlTakeuchi2008-HsnbgLmt, XiangHofmannPryde2013-MltPhPhsSens}.
  On the other hand, the estimation resolution in average photon number estimation is relatively robust against photon losses.
  In this section, we show that even small photon losses can remove the contributions of destructive quantum interferences from the Fisher information obtained in PNRD measurements.
  We will employ this result to develop a theory of quantification of contributions of destructive quantum interferences in the quantum Fisher information, which can be used to describe the effects of losses in PNRD measurements.

\label{sec::reduction_FI_in_lossy_PNRD}
  \begin{figure}[t]
    \includegraphics[width=0.48\textwidth]{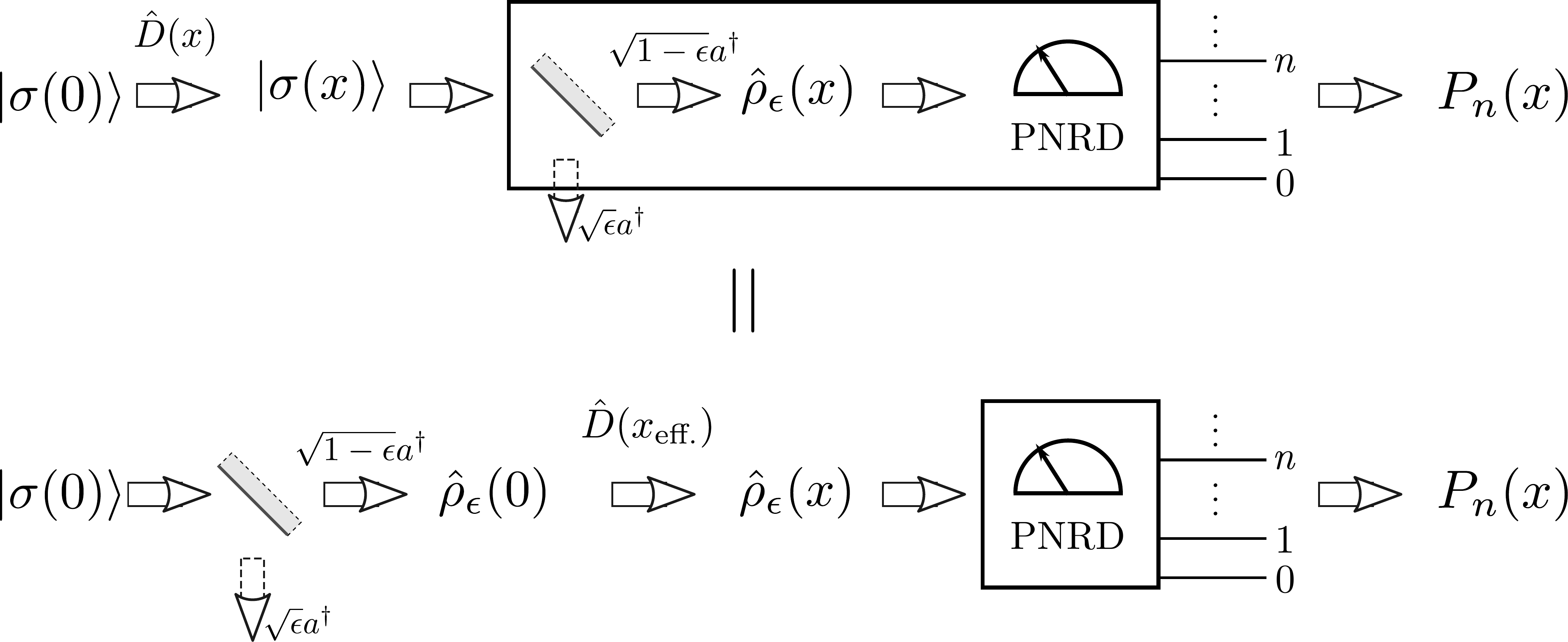}
    \caption{A lossy PNRD measurement of $x$-displaced squeezed states $\ket{\sigma(x)}$ with the detector efficiency $(1-\epsilon)$ (the upper part).
    The model of photon losses is described by a beam splitter with the reflection coefficient $\epsilon$.
    The input pure state $\ket{\sigma(x)}$ undergoing photon losses becomes a mixed state $\hat{\rho}_{\epsilon}(x)$.
    The resolved photon number distribution $P_{n}(x)$ given in Eq. \eqref{eq::prob_lossy_PNRD} is associated with the mixed state $\hat{\rho}_{\epsilon}(x)$.
    The lossy PNRD measurement in the upper part is equivalent to the lower part, in which the initial squeezed vacuum state $\ket{\sigma(0)}$ first passes through the lossy channel (the beam splitter), then is displaced by the displacement operator $\hat{D}(x_{\text{eff.}})$ with an effective displacement $x_{\text{eff.}}=\sqrt{1-\epsilon}\;x$.
    In the end, the final state $\rho_{\epsilon}(x)$ (see Eq. \eqref{eq::lossy_displ_sq_st}) is resolved by a perfect PNRD.
    }
    \label{fig::scheme_lossy_PNRD}
  \end{figure}

  In a lossy PNRD measurement, the efficiency $(1-\epsilon)$ of a PRND is characterized by the rate of photon losses $\epsilon$, which is the probability of a single photon being lost in the detector.
  The model of photon losses in a PNRD can be described by a beam splitter with the reflection coefficient $\epsilon$ (see Fig. \ref{fig::scheme_lossy_PNRD}).
  The photon number distribution $P_{n}$ resolved in a lossy PNRD measurement is given by the sum of the probabilities of the $(n+k)$-photon number inputs losing $k$ photons in the outputs,
  \begin{equation}
  \label{eq::prob_lossy_PNRD}
    P_{n}(\epsilon, x)
    =
    (1-\epsilon)^{n}\sum_{k}\binom{n+k}{k}\epsilon^{k}p_{n+k}(x),
  \end{equation}
  where $p_{n+k}(x) = |\braket{n+k|\sigma(x)}|^{2}$ is the photon number distribution of the pure state inputs.
  The Fisher information obtained in such a lossy PNRD measurement is determined by
  the logarithmic derivatives $\partial_{x}(\ln P_{n} )$ where the contribution $I_{n}$ of each outcome $n$ is given by
  \begin{align}
  \label{eq::n_ph_contr}
    I_{n}(\epsilon,x)
    = & \frac{\left(\partial_{x}P_{n}(\epsilon,x)\right)^{2}}{P_{n}(\epsilon,x)}.
  \end{align}
  The total Fisher information $I_{F}$ is then given by the sum of the contributions $I_{n}$ from all the $n$-photon outputs
  \begin{equation}
    I_{F}(\epsilon, x)
    =
    \sum_{n}I_{n}(\epsilon,x).
  \end{equation}
  The amount of change in the Fisher information by photon losses depends critically on both the magnitudes of the probabilities and their derivatives.
  Specifically, the derivatives will be zero at maxima and minima of the $x$-dependence of the probability $P_{n}(\epsilon,x)$.
  Equation \eqref{eq::n_ph_contr} indicates that the Fisher information $I_{n}$ contributed by these results will then go to zero unless the probability itself also goes to zero.
  In the presence of small losses, the minimal probabilities at $x_{n,k}$ will be greater than zero ($P_{n}(\epsilon,x_{n,k})>0$), so that the contributions to the Fisher information for these minima are exactly zero,
  \begin{equation}
  \label{eq::zero_I_for_non-zero_min_1}
    I_{n}(\epsilon>0, x_{n,k}) = 0.
  \end{equation}
  However, the original probabilities of the pure state minima are all zero $(p_{n}(x_{n,k})=0)$.
  This means that they contribute a finite amount of Fisher information, given by the second derivative of the pure state probability,
  \begin{equation}
  \label{eq::zero_I_for_non-zero_min}
    I_{n}(\epsilon=0, x_{n,k}) = 2\partial_{x}^{2}p_{n}(x_{n,k}).
  \end{equation}
  It should be noted that this contribution to the Fisher information is essential for achieving the QCR bound in the presence of quantum interferences.
  Even a small rate of photon losses $\epsilon\ll1$ will immediately reduce the Fisher information at the minimum to zero.
  However, the precise $x$-dependence of this loss of Fisher information requires a more detailed analysis of the relation between the pure state probabilities $|\braket{n|\sigma(x)}|^{2}$ and the detection probabilities $P_{n}(\epsilon, x)$ in the presence of losses.

  \bigskip

  The detection probabilities $P_{n}(\epsilon, x)$ for a lossy PNRD can be described by applying a linear optics loss rate of $\epsilon$ to the squeezed vacuum state before applying displacement operation $\hat{D}(x_{\text{eff.}})$, in which the effective displacement $x_{\text{eff.}}=\sqrt{1-\epsilon}x$ has also been modified by the losses (see Fig. \ref{fig::scheme_lossy_PNRD}).
  The detection probabilities of the lossy PNRD are then described by the photon number distribution $P_{n}$ of a mixed state $\rho_{\epsilon}(x)$,
  \begin{equation}
    P_{n}(\epsilon, x) = \braket{n|\hat{\rho}_{\epsilon}(x)|n},
  \end{equation}
  where the effects of losses have been included in the $\epsilon$-dependence of the mixed state.
  It is possible to separate the displacement from the main effect of the losses using
  \begin{equation}
  \label{eq::lossy_displ_sq_st}
    \hat{\rho}_{\epsilon}(x)
    =
    \hat{D}(x_{\text{eff.}})
    \hat{\rho}_{\epsilon}(0)
    \hat{D}^{\dagger}(x_{\text{eff.}}).
  \end{equation}
  Here, $\hat{\rho}_{\epsilon}(0)$ is the squeezed thermal state that results when photon losses are applied to a squeezed vacuum\cite{MarianMarian1993-SqStThmlNoiseI,MarianMarian1993-SqStThmlNoiseII,ParisEtAlSiena2003-PurityGaussStInNsyChnnl,SerafiniEtAlSiena2005-QDecohCVSys,MonrasParis2007-QEstOfLoss,OnoHofmann2010-PhEst},
  \begin{align}
    \hat{\rho}_{\epsilon}(0)
    & =
    (1-\lambda)\sum_{m}\lambda^{m}\projector{\Phi_{m}},
  \end{align}
  where the $\ket{\Phi_{m}}$ are squeezed photon number states and the thermal coefficient $\lambda$ is given by
  \begin{equation}
    \lambda  =
    \frac{\sqrt{1+4(1-\epsilon)\epsilon\sinh^{2}(r)}-1}{\sqrt{1+4(1-\epsilon)\epsilon\sinh^{2}(r)}+1}.
  \end{equation}
  The squeezed photon number states can be described by a unitary squeezing operator acting on an $m$-photon state
  \begin{equation}
    \ket{\Phi_{m}}
    =
    \hat{S}(r_{\text{eff.}})\ket{m},
  \end{equation}
  where the effective squeezing parameter $r_{\text{eff.}}$ is a function of the original squeezing parameter $r$  and the loss rate $\epsilon$,
  \begin{align}
  \label{eq::eff_sq_factor}
    r_{\text{eff.}} =
    \frac{1}{4}\ln\left(\frac{(1-\epsilon)e^{2r}+\epsilon}{(1-\epsilon)e^{-2r}+\epsilon}\right).
  \end{align}
  For small photon losses $\epsilon\ll1$, the thermal coefficient $\lambda$ is approximately equal to $\epsilon\sinh^{2}(r)\ll1$ and the mixed state $\rho_{\epsilon}(0)$ can be approximated by a mixture of squeezed vacuum and squeezed single photon state,
  \begin{align}
  \label{eq::mix_rho_0}
    \hat{\rho}_{\epsilon}(0)
    \approx &
    \left(1-\epsilon\sinh^{2}(r)\right)\projector{\Phi_{0}}
    \nonumber \\
    & + \epsilon\sinh^{2}(r)\projector{\Phi_{1}}.
  \end{align}
  The contribution of the displaced squeezed vacuum $\ket{\Phi_{0}(x_{\text{eff.}})} = \hat{D}(x_{\text{eff.}})\ket{\Phi_{0}}$ corresponds to the probability distribution $p_{n}(x_{\text{eff.}})$ of the pure state input $\ket{\sigma(x_{\text{eff.}})}$ with a squeezing parameter of $r_{\text{eff.}}$.
  Therefore the probability $P_{n}(\epsilon, x)$ is very close to $p_{n}(x_{\text{eff.}})$ for a squeezing parameter of $r_{\text{eff.}}$.

  To estimate the effect of small losses, we can now focus on the zero points in the distribution $p_{n}(x_{\text{eff.}})$ defined by the $x_{\text{eff.}}$-displaced $r_{\text{eff.}}$-squeezed vacuum contribution in Eq. \eqref{eq::mix_rho_0},
  \begin{equation}
    x_{\text{eff.}} = x_{n,k}.
  \end{equation}
  Around these points, the contribution of the squeezed single photon state should be added to the probability resulting in a modified Fisher information of
  \begin{align}
  \label{eq::In_apprx_tmp_1}
    & I_{n}(\epsilon, x_{\text{eff.}})
    \approx
    \frac{
      (\partial_{x_{\text{eff.}}}p_{n}(x_{\text{eff.}}))^{2}
    }{
      p_{n}(x_{\text{eff.}}) + \epsilon \sinh^{2}(r) |\braket{n|\Phi_{1}(x_{\text{eff.}})}|^{2}
    },
  \end{align}
  where $\ket{\Phi_{1}(x_{\text{eff.}})}$ is the $r_{\text{eff.}}$-squeezed single photon state displaced by $x_{\text{eff.}}$.
  Since we can assume that the probability contributed by this state varies very little with $x$, it is possible to replace it with its value at $x_{\text{eff.}}=x_{n,k}$.
  Using the Taylor expansion of $p_{n}(x_{\text{eff.}})$, it is possible to relate the Fisher information in Eq. \eqref{eq::In_apprx_tmp_1} to the Fisher information $\mathcal{I}_{n}(x_{\text{eff.}})$ obtained with perfect photon detection for a squeezing parameter $r_{\text{eff.}}$ and a displacement of $x_{\text{eff.}}$.
  The result can expressed by a reduction factor $\delta_{n}$ defined as
  \begin{equation}
    \delta_{n} = 1-\frac{I_{n}}{\mathcal{I}_{n}}
    \text{\;\; with \;\;}
    \mathcal{I}_{n} = \frac{(\partial_{x_{\text{eff.}}}p_{n}(x_{\text{eff.}}))^{2}}{p_{n}(x_{\text{eff.}})}.
  \end{equation}
  The displacement dependence of the reduction factor in the vicinity of a single zero point $x_{n,k}$ is given by
  \begin{equation}
  \label{eq::delta_n_form}
    1-\delta_{n}(\epsilon, x_{\text{eff.}})
    =
    \frac{
      (x_{\text{eff.}}-x_{n,k})^{2}
    }{
      (x_{\text{eff.}}-x_{n,k})^{2} + \epsilon\beta_{n,k}
    },
  \end{equation}
  where the sharpness coefficient $\beta_{n,k}$ is given by
  \begin{equation}
    \beta_{n,k}
    =
    \sinh^{2}(r)
    \left.
      \frac{
        |\braket{n|\Phi_{1}(z)}|^{2}
      }{
        \partial_{z}^{2}
        p_{n}(z)
      }
    \right|_{z=x_{n,k}}.
  \end{equation}
  It is possible to solve this equation using the photon number distributions of the displaced squeezed number states $\ket{\Phi_{0}}$ and $\ket{\Phi_{1}}$.
  The result does not depend on the location of the minimum given by $(n,k)$.
  It is instead described by a constant value determined by the squeezing parameters $r$ and $r_{\text{eff.}}$,
  \begin{equation}
    \beta_{n,k} = \beta = \sinh^{2}(r)e^{-2r_{\text{eff.}}}.
  \end{equation}
  Since the sigmoid function in Eq. \eqref{eq::delta_n_form} may have a non-trivial overlap with functions obtained from neighboring minima, it is useful to interpolate the reduction factor between the minima by a product,
  \begin{equation}
  \label{eq::delta_n_form_general}
    1-\delta_{n}(\epsilon, x_{\text{eff.}})
    =
    \prod_{k}
    \frac{
      (x_{\text{eff.}}-x_{n,k})^{2}
    }{
      (x_{\text{eff.}}-x_{n,k})^{2} + \epsilon\beta
    }.
  \end{equation}
  It is therefore possible to describe the reduction of Fisher information by small photon losses using reduction factors determined mostly by the locations $x_{n,k}$ of destructive interferences in the photon number distribution of displaced squeezed vacuum states.

  \begin{figure*}[t]
    \centering%
    \hfill%
    \subfloat[]{
        \includegraphics[width=0.48\textwidth]{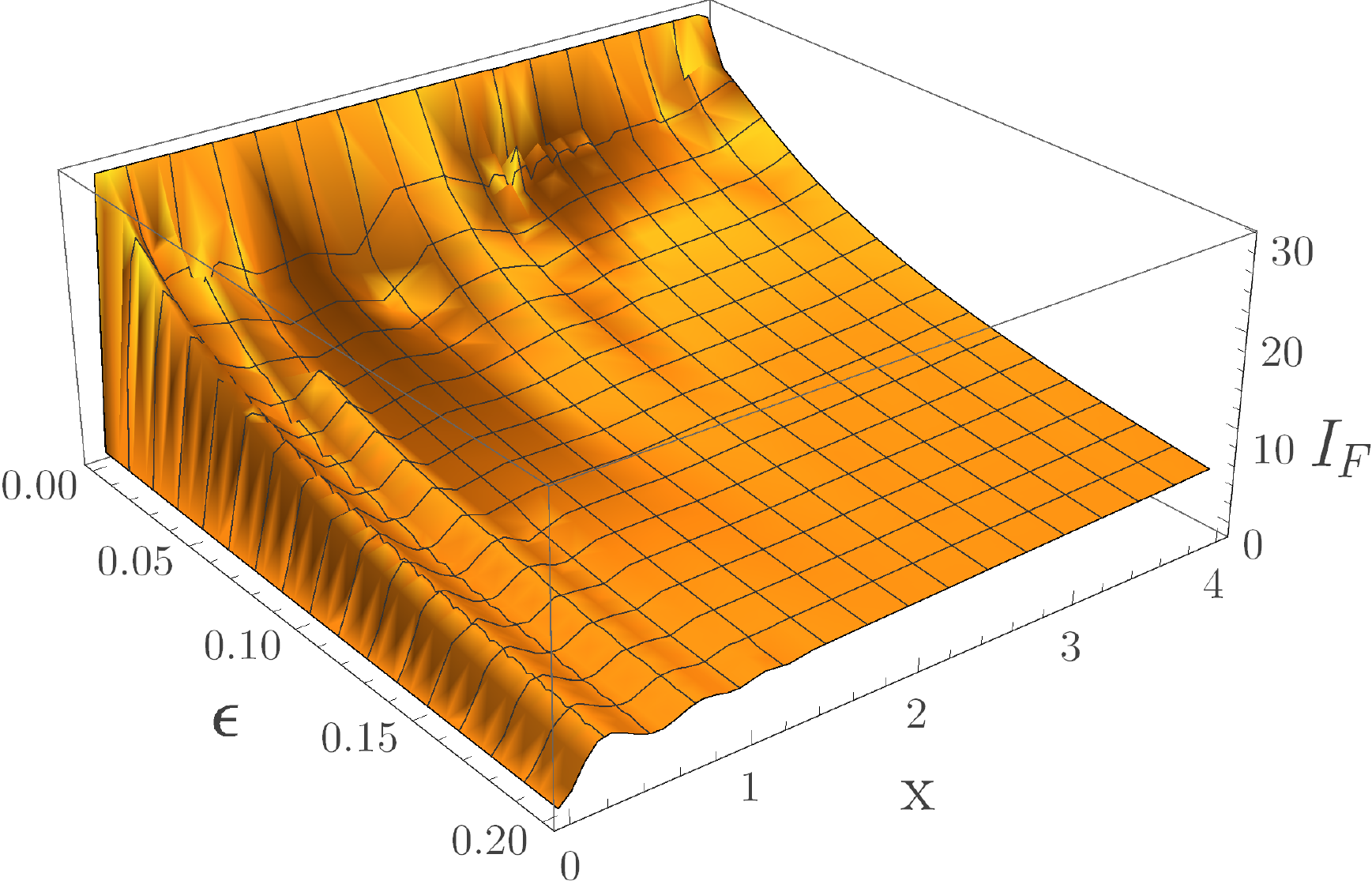}
    }
    \hfill%
    \subfloat[]{
        \includegraphics[width=0.48\textwidth]{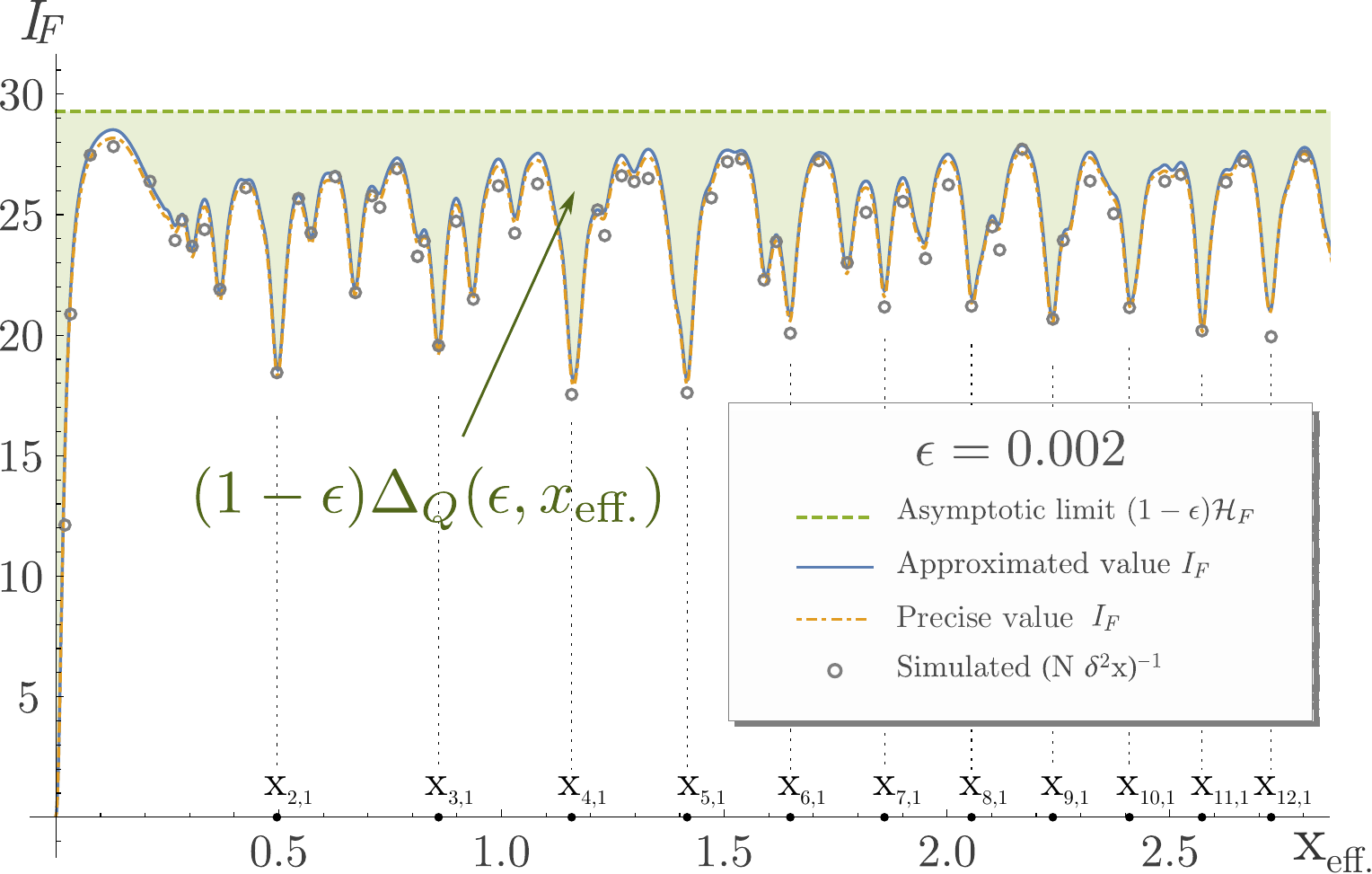}
    }
    \hfill%
    \hfill%
    \hfill%
    \\
    \hfill%
    \subfloat[]{
        \includegraphics[width=0.48\textwidth]{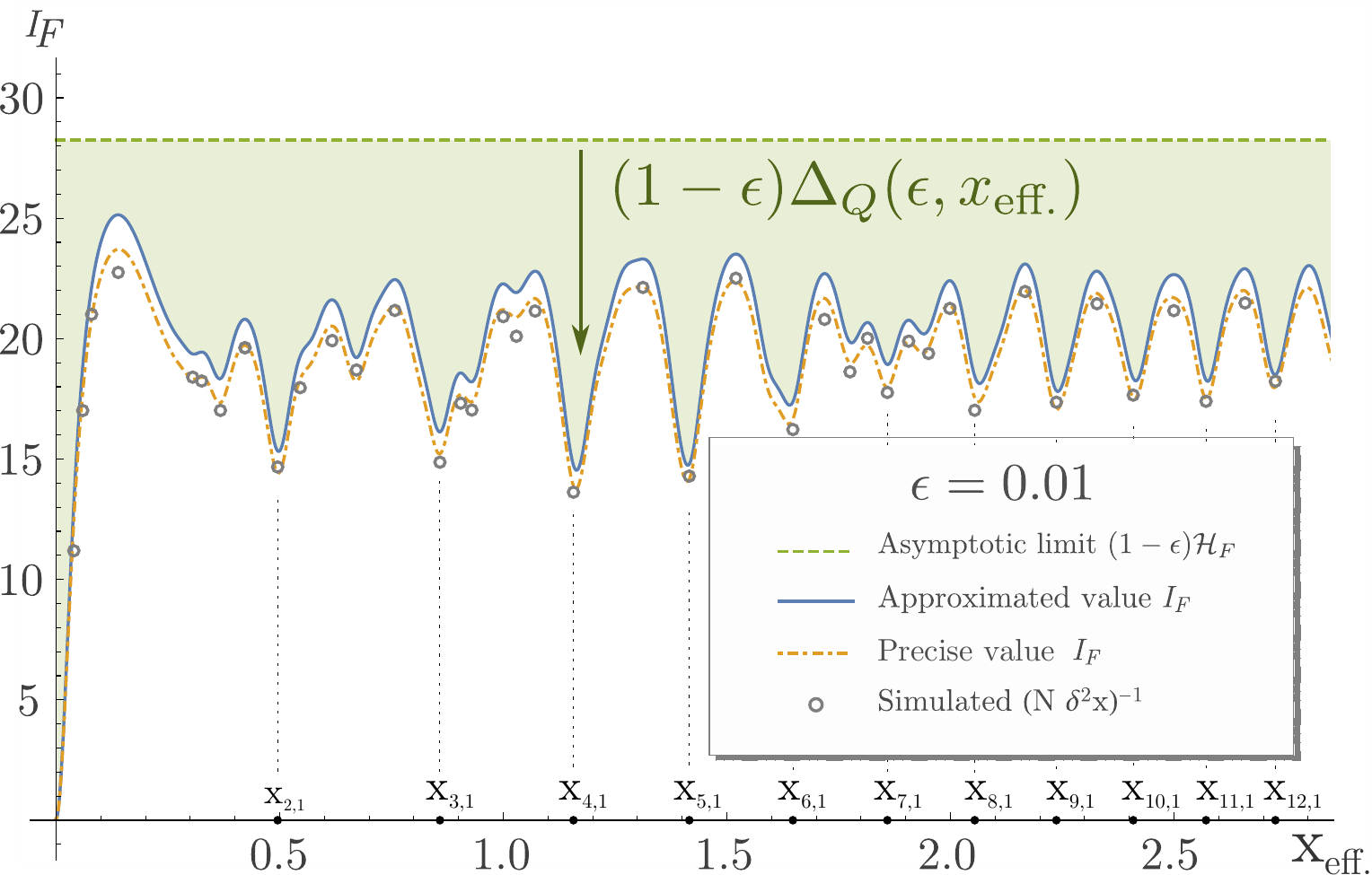}
    }
    \hfill%
    \subfloat[]{
        \includegraphics[width=0.48\textwidth]{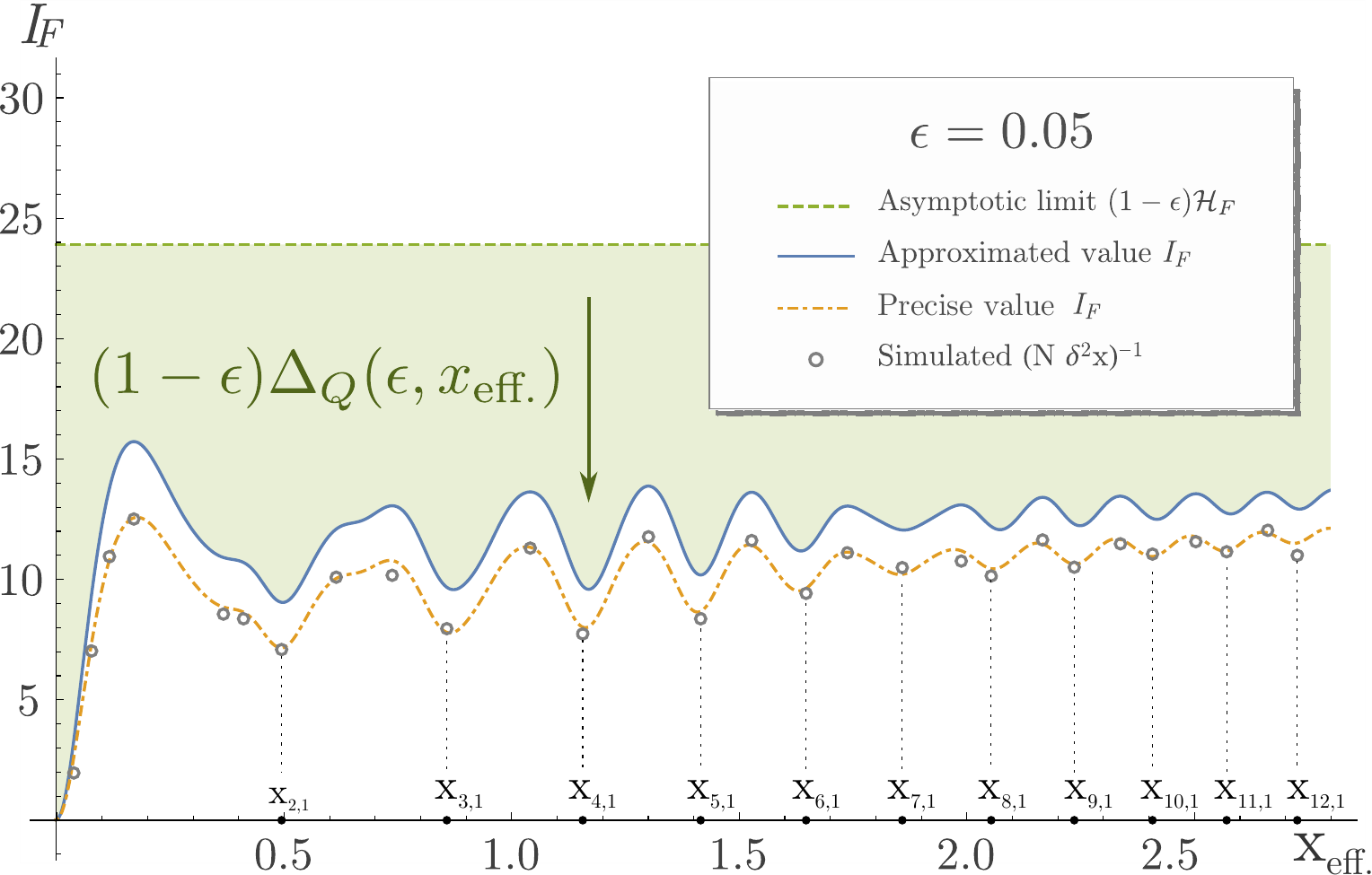}
    }
    \hfill%
    \hfill%
    \caption{\colorfig
      Fisher information $I_F$ of the $x$-estimation of the displaced squeezed state $\ket{\sigma(x)}$ in lossy PNRD measurements with photon losses $\epsilon$ and a squeezing parameter of $r=1$.
      The approximated Fisher information is given in Eq. \eqref{eq::FI_PNR_reduced}.
      Panel (a) shows the $x$- and $\epsilon$-dependence of the approximated Fisher information $I_F$.
      Panel (b) shows the $x_{\text{eff.}}$-dependence of the Fisher information $I_F$ for a small photon loss $\epsilon = 0.002$.
      The green dashed line $(1-\epsilon)\mathcal{H}_{F}$ is the asymptotic limit of the approximated Fisher information $I_{F}$ (the blue solid line) for $x\to \infty$.
      The approximated Fisher information is very close to the value obtained from the precise photon number distributions (the orange dot-dashed line).
      The green highlighted area between the asymptotic limit and the approximated Fisher information is the reduction function $\Delta_{Q}$ given in Eq. \eqref{eq::reduction_fct_sq_st}.
      The white circles are the estimation sensitivity $1/(N\delta^{2} x)$ obtained from a numerical simulation of a lossy PNRD estimation using $N=2000$ samples.
      Panel (c) shows the $x_{\text{eff.}}$-dependence of the Fisher information for photon losses of $\epsilon = 0.01$. As the photon losses increase, the approximation becomes less accurate, and the structure of the dips is broadened.
      Panel (d) shows the $x_{\text{eff.}}$-dependence of the Fisher information for photon losses of $\epsilon = 0.05$.  The main dips now appear as a small modulation of a nearly homogeneous reduction of Fisher information.
    }\label{fig::q_interf_in_FI}
  \end{figure*}
\begin{figure*}[t]
  \centering
  \subfloat[]{\includegraphics[width=0.45\textwidth]{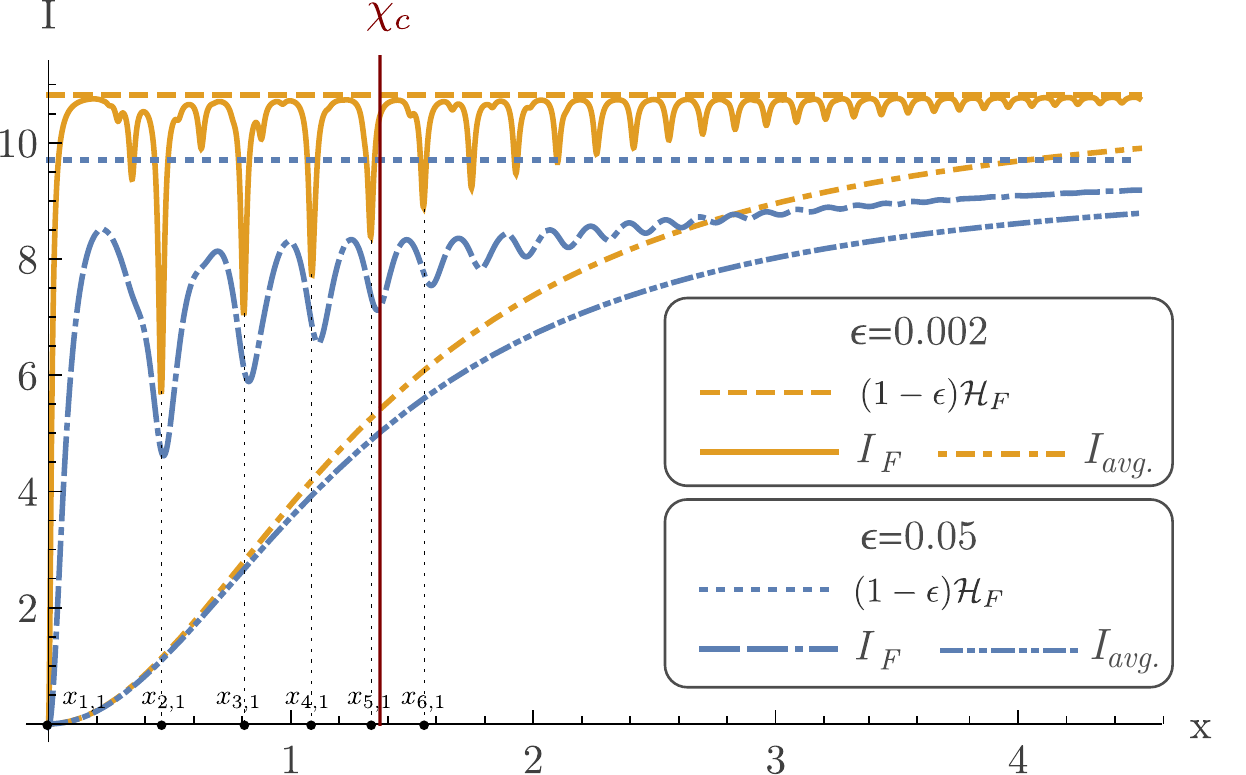}}
  \hfill
  \subfloat[]{\includegraphics[width=0.45\textwidth]{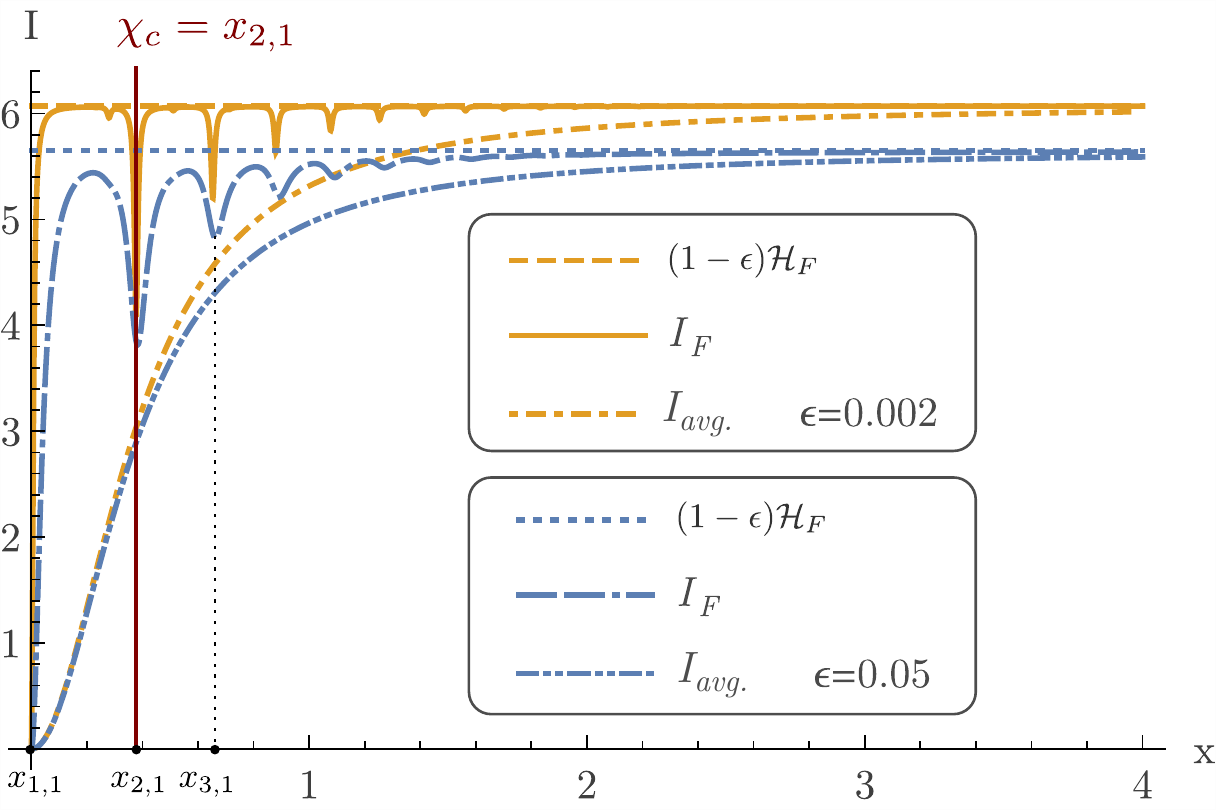}}
  \hfill
  \caption{
    Comparison of Fisher information $I_{F}$ of lossy PNRD estimation with the sensitivity $I_{\text{avg.}}$ of average photon number estimates.
    The orange lines show photon losses of $\epsilon = 0.002$ and the blue lines show photon losses of $\epsilon = 0.05$.
    The asymptotic limits $(1-\epsilon)\mathcal{H}_{F}$ are shown by the orange dashed line and the blue dotted line, respectively.
    The sensitivity of the average photon number estimate $I_{\text{avg.}}$ are shown by the orange dash-dotted line and the blue dash-double-dotted line, respectively.
    Photon losses reduce the Fisher information to a value between the asymptotic limit for $x\to\infty$ and the sensitivity for the average photon number estimate.
    Panel (a) shows the Fisher information for an input state with a squeezing parameter of $r=0.5$.
    The critical displacement $\chi_{c}\approx1.37$ is above the zero point $x_{5,1}$. One can observe significant contributions of destructive quantum interferences at $x_{n,1}$ for $n\le5$.
    Panel (b) shows the Fisher information for an input state with a squeezing parameter of $r=0.21$. The critical displacement $\chi_{c}\approx0.38$ is significantly lower. There are no additional dips between $x_{1,1}=0$ and $x_{2,1}=\chi_{c}$.
  }
  \label{fig::asympt_behavior}
\end{figure*}

  Summing up all the $n$-photon contributions $I_{n}$ in Eq. \eqref{eq::In_apprx_tmp_1}, the total Fisher information is approximately given by the quantum Fisher information $\mathcal{H}_{F}(\epsilon)$ of the pure state $\ket{\Phi_{0}(x_{\text{eff.}})}$ minus a reduction term $\Delta_{Q}(\epsilon,x_{\text{eff.}})$,
  \begin{align}%
  \label{eq::FI_PNR_reduced}
    I_{F}(\epsilon,x)
    & \approx 
    (1-\epsilon)
    \left(\mathcal{H}_{F}(\epsilon)-\Delta_{Q}\left(\epsilon, x_{\text{eff.}}\right)\right).
  \end{align}%
  The global reduction factor of $(1-\epsilon)$ is a result of the re-scaling of the displacement from $x_{\text{eff.}}$ to $x$.
  The quantum Fisher information $\mathcal{H}_{F}(\epsilon)$ does not depend on displacement $x$ and is given by
  \begin{equation}
  \label{eq::asymp_IFisher}
    \mathcal{H}_{F}(\epsilon)
    =
    4\sqrt{\frac{(1-\epsilon)e^{2r}+\epsilon}{(1-\epsilon)e^{-2r}+\epsilon}}.
  \end{equation}
  The complete reduction term is a sum of the reductions in Fisher information for each photon number $n$,
  \begin{equation}
  \label{eq::reduction_fct_sq_st}
    \Delta_{Q}(\epsilon,x_{\text{eff.}})
    =
    \sum_{n}
    \mathcal{I}_{n}(\epsilon,x_{\text{eff.}})
    \delta_{n}(\epsilon,x_{\text{eff.}}).
  \end{equation}
  The reduction function $\Delta_{Q}$ is a characteristic function of destructive quantum interferences in $\braket{n|\Phi_{0}(x)}$.
  If no photon is lost in the PNRD measurement, $\Delta_{Q}(0,x)$ is zero and $\mathcal{H}_{F}(0)$ is the quantum Fisher information of the original $x$-displaced $r$-squeezed state $\ket{\sigma(x)}$ given in Eq. \eqref{eq::sq_st}.
  The reduction in Fisher information caused by photon losses are described by the sigmoid function given in Eqs. \eqref{eq::delta_n_form} and \ref{eq::delta_n_form_general}.
  The sharpness of the dips described by these functions is determined by the product of the sharpness coefficient and the photon loss rate $\epsilon$.
  For very small photon losses, the reduction term $\Delta_{Q}$ describes a series of well separated sharp dips in Fisher information at the zero points $x_{n,k}$ of the input pure state.
  The precise displacement of the dips is given by $x = x_{n,k}/\sqrt{1-\epsilon}$.
  The depth of the dips is given by
  \begin{equation}
  \label{eq::reduction_peak_n_ph_contr}
    \Delta_{Q}(\epsilon, x_{n,k})
    =
    \mathcal{I}_{n}(\epsilon, x_{n,k}).
  \end{equation}
  As displacement increases, the values $\mathcal{I}_{n}(\epsilon, x_{n,k})$ decrease because the zero points $x_{n,k}$ occur outside of main peak of the probability distribution.
  As a result, the reduction function asymptotically converges to zero as $x$ increases, and $I_{F}(\epsilon,x)$ has an asymptotical limit of
  \begin{equation}
    I_{F}(\epsilon, x\rightarrow\infty)
    =
    (1-\epsilon)\mathcal{H}_{F}(\epsilon).
  \end{equation}
  The smaller the photon loss is, the sharper the reduction dips are, and hence the structure of destructive quantum interferences can be observed directly in the Fisher information extracted from the photon statistics.
  Interestingly, the transition to a perfect PRND given $\epsilon\to 0$ leaves the depth of the dips constant, making them disappear at $\epsilon=0$ only as a consequence of their sharpness.
  As long as the photon losses are nonzero, the positions of the zero points $x_{n,k}/\sqrt{1-\epsilon}$ are clearly visible in the displacement dependence of Fisher information.

  Figure \ref{fig::q_interf_in_FI} shows the dependence of Fisher information on the displacement for various photon loss rate $\epsilon$.
  For the small photon loss of $\epsilon=0.002$ shown in Fig. \ref{fig::q_interf_in_FI}(b),  the approximated value of $I_{F}$ obtained from Eq. \eqref{eq::FI_PNR_reduced} is sufficiently accurate.
  The Fisher information is reduced from the asymptotic limit $\mathcal{H}_{F}$ with a series of main dips at the zero points $x_{n,1}$ given in Eq. \eqref{eq::zero_sq_st_exact}.
  Since the photon loss is small, the reduction dips are sufficiently sharp to observe several subdips belonging to other zero points, such as $x_{n,2}$.
  At $x=0$, where all probabilities are either  at a minimum or a maximum,  the Fisher information is reduced to zero by the losses.
  At a larger photon loss of $\epsilon = 0.01$ shown in Fig. \ref{fig::q_interf_in_FI}(c), the maximal Fisher information obtained between the dips is significantly lower than the asymptotical limit of Fisher information at $x\to\infty$.
  The reason is that the reduction dips overlap significantly.
  As a result, the sub-dips disappear and the remaining structure of the displacement dependence is mostly associated with the position of the minima $x_{n,1}$.
  For an even larger photon loss of $\epsilon = 0.05$ shown in Fig. \ref{fig::q_interf_in_FI} (d), the distinct dips have nearly disappeared, leaving a slightly modulated reduction of Fisher information to about half of the asymptotic value.
  These results implicate that the bias phase $\varphi$ in two-path interferometers should be set between two neighboring zero points $(2 x_{n,1}/\alpha)$ and $(2 x_{n+1,1}/\alpha)$ for achieving high sensitivity in lossy PNRD-estimation.

  \bigskip


  The reduction dips show that the Fisher information obtained from the complete photon number distribution resolved in PNRDs is very sensitive to photon losses.
  Here, we want to compare the fragility of the Fisher information $I_{F}(\epsilon,x)$ under photon losses with the estimation sensitivity $I_{\text{avg.}}$ of average number estimates.
  The explicit sensitivity $I_{\text{avg.}}$ can be determined by Eq. \eqref{eq::I_cl_derivation} and is approximately given by the following equation for small photon losses:
  \begin{equation}
  \label{eq::avg_num_res_e}
    I_{\text{avg.}}(\epsilon, x)
    \underset{\epsilon\ll1}{\approx}
    (1-\epsilon)I_{\text{avg.}}(0, x),
  \end{equation}
  where $I_{\text{avg.}}(0, x)$ is the sensitivity of the average number estimation of the $x$-displaced $r$-squeezed state $\ket{\sigma(x)}$ given in Eq. \eqref{eq::I_cl}.
  Since the estimation using a complete photon number distribution $P_{n}(x)$ should always be better than the average photon number estimation, the reduction of Fisher information induced by photon losses should always be bounded
  by the sensitivity $I_{\text{avg.}}(\epsilon,x)$ obtained in the average photon number estimation,
  \begin{equation}
    I_{\text{avg.}}(\epsilon,x)
    \le
    I_{F}(\epsilon,x).
  \end{equation}
  According to Eq. \eqref{eq::I_cl}, the average photon number resolution exhibits a transition approaching the quantum Fisher information at high displacements.
  This should also put a limit on the depth of reduction dips as displacement increases.

  Figure \ref{fig::asympt_behavior} shows the Fisher information for squeezing parameters of $r=0.5$ and $r=0.21$.
  For $r=0.5$, one can observe significant reductions of Fisher information at the first five zero-probability points $x_{n,1}$, which quantify the contributions of the destructive quantum interferences in $I_{F}$ at these points.
  The depth of the reduction dips decrease with increasing displacement, in parallel to the transition of the sensitivity $I_{\text{avg.}}$ of the average photon number estimation.
  A similar behavior is observed at $r=0.21$, where the first dip after $x=0$ occurs at the critical displacement $\chi_{c}=x_{2,1}$.
  The depth of the dip is clearly limited by the sensitivity of the average photon number estimation.

\section{Conclusion}
\label{sec::conclusion}

In this paper, we have shown that photon-number-resolving detections (PNRDs) can be employed in the dark output port of a two-path interferometer operating at high intensity to extract the quantum Fisher information of a small phase shift.
In such interferometric settings, the phase estimation is approximately equivalent to the $x$-quadrature displacement estimation on the displaced states in the dark output port.
For displaced squeezed states, there is a transition point $\chi_{c}$ for the $x$-quadrature displacement, upon which the average photon number estimate approaches the QCR bound.
The underlying reason for such a transition has been shown to be the
transition of photon statistics of displaced squeezed states from quantum interferences to a single-peaked Gaussian distribution.
For the $x$-quadratures displacement below the critical point, the quantum Fisher information is significantly contributed by quantum interference in photon number statistics, which is sensitive to photon losses in PNRDs.

The dark port regime, which is defined by the critical phase corresponding to $\chi_{c}$, is characterized by quantum interference patterns that are quite similar to the multiphoton interferences which are usually observed in highly nonclassical states such as NOON or Holland-Burnett states.
It is interesting that the same characteristic can also be observed in high-intensity two-path interferometers when the quantum enhancement is achieved by squeezing the vacuum in the dark input port.
The complete phase sensitivity of the undetected photons in the bright output port can then be transferred to very few photons exhibiting genuine multiphoton interference patterns in the dark output port.
As we have demonstrated here, it is possible to access the Fisher information in the dark output port by photon-number-resolving detection of the few photons in that port, if the detection losses are sufficiently small.
The pattern of Fisher information reduction caused by small photon losses is highly characteristic of the quantum interference fringes that characterize nonclassical light.
It is therefore possible to identify a genuine quantum advantage in the dark port detection of the interference between squeezed vacuum and high-intensity coherence light.
It seems to be remarkable that this advantage allows us to effectively compress the Fisher information of a huge photon number into the Fisher information obtained from a precise detection of very few photons.
We think that, aside from the possible practical advantages, this result has interesting implications for the relation between the suppression of quantum fluctuations by squeezing and the multiphoton interference fringes observed in NOON states or other highly nonclassical superposition states.

\bigskip

\begin{acknowledgments}
  This work is supported by JST-CREST (JPMJCR1674), Japan Science and Technology Agency.
\end{acknowledgments}

\bibliographystyle{myunsrt}
\bibliography{QFI_under_ph_losses}

\begin{thebibliography}{10}

\bibitem{LIGO2011-GW}
{The LIGO Scientific Collaboration}.
\newblock A gravitational wave observatory operating beyond the quantum
  shot-noise limit.
\newblock {\em Nature Physics}, 7:962, 2011.

\bibitem{LIGO2013-GW}
{The LIGO Scientific Collaboration}.
\newblock Enhanced sensitivity of the ligo gravitational wave detector by using
  squeezed states of light.
\newblock {\em Nature Photonics}, 7:613, 2013.

\bibitem{LIGO2016-GW}
{The LIGO Scientific Collaboration and Virgo Collaboration}.
\newblock Observation of gravitational waves from a binary black hole merger.
\newblock {\em Phys. Rev. Lett.}, 116:061102, 2016.

\bibitem{Caves1981-QNoiseSqSt}
C.~M. Caves.
\newblock Quantum-mechanical noise in an interferometer.
\newblock {\em Phys. Rev. D}, 23:1693--1708, 1981.

\bibitem{LangCaves2013-OptQEnhance}
M.~D. Lang and C.~M. Caves.
\newblock Optimal quantum-enhanced interferometry using a laser power source.
\newblock {\em Phys. Rev. Lett.}, 111:173601, 2013.

\bibitem{YurkeMcCallKlauder1986-SU2SU11Interf}
B.~Yurke, S.~L. McCall, and J.~R. Klauder.
\newblock {SU}(2) and {SU}(1,1) interferometers.
\newblock {\em Phys. Rev. A}, 33:4033--4054, 1986.

\bibitem{HilleryMlodinow1993-InterfNSqSt}
M.~Hillery and L.~Mlodinow.
\newblock Interferometers and minimum-uncertainty states.
\newblock {\em Phys. Rev. A}, 48:1548--1558, 1993.

\bibitem{Paris1995-CohSqNSqSqInterf}
M.~G. Paris.
\newblock Small amount of squeezing in high-sensitive realistic interferometry.
\newblock {\em Physics Letters A}, 201(2):132 -- 138, 1995.

\bibitem{BarnettMaitre2003}
{Barnett, S. M.}, {Fabre, C.}, and {Ma\^{\i}tre, A.}
\newblock Ultimate quantum limits for resolution of beam displacements.
\newblock {\em Eur. Phys. J. D}, 22(3):513--519, 2003.

\bibitem{AtamanPredaIonicioiu2018-PhSnsng1MdN2Md}
S.~Ataman, A.~Preda, and R.~Ionicioiu.
\newblock Phase sensitivity of a {Mach-Zehnder} interferometer with
  single-intensity and difference-intensity detection.
\newblock {\em Phys. Rev. A}, 98:043856, 2018.

\bibitem{BraunsteinCaves1994-StDistGeoQSt}
S.~L. Braunstein and C.~M. Caves.
\newblock Statistical distance and the geometry of quantum states.
\newblock {\em Phys. Rev. Lett.}, 72:3439--3443, 1994.

\bibitem{PezzeSmerzi2008-MZIHsnbgLmtCSV}
L.~Pezz\'e and A.~Smerzi.
\newblock {Mach-Zehnder} interferometry at the {Heisenberg} limit with coherent
  and squeezed-vacuum light.
\newblock {\em Phys. Rev. Lett.}, 100:073601, 2008.

\bibitem{Hofmann2009-PureStHsnbgLmt}
H.~F. Hofmann.
\newblock All path-symmetric pure states achieve their maximal phase
  sensitivity in conventional two-path interferometry.
\newblock {\em Phys. Rev. A}, 79:033822, 2009.

\bibitem{OnoHofmann2010-PhEst}
T.~Ono and H.~F. Hofmann.
\newblock Effects of photon losses on phase estimation near the {Heisenberg}
  limit using coherent light and squeezed vacuum.
\newblock {\em Phys. Rev. A}, 81:033819, 2010.

\bibitem{SeshadreesanEtAlDowling2011-ParityDtctnHsnbgLmt}
K.~P. Seshadreesan, P.~M. Anisimov, H.~Lee, and J.~P. Dowling.
\newblock Parity detection achieves the {Heisenberg} limit in interferometry
  with coherent mixed with squeezed vacuum light.
\newblock {\em New Journal of Physics}, 13(8):083026, 2011.

\bibitem{MillerEtAlSergienko2003-PNRD}
A.~J. Miller, S.~W. Nam, J.~M. Martinis, and A.~V. Sergienko.
\newblock Demonstration of a low-noise near-infrared photon counter with
  multiphoton discrimination.
\newblock {\em Applied Physics Letters}, 83(4):791--793, 2003.

\bibitem{KhouryEtBouwmeester2006-PNRDOfNnlnrInterf}
G.~Khoury, H.~S. Eisenberg, E.~J.~S. Fonseca, and D.~Bouwmeester.
\newblock Nonlinear interferometry via {Fock}-state projection.
\newblock {\em Phys. Rev. Lett.}, 96:203601, 2006.

\bibitem{LitaMillerNam2008-PNRD}
A.~E. Lita, A.~J. Miller, and S.~W. Nam.
\newblock Counting near-infrared single-photons with 95\% efficiency.
\newblock {\em Optics express}, 16(5):3032--3040, 2008.

\bibitem{WildfeuerEtAlDowling2009-PNRDFPInterf}
C.~F. Wildfeuer, A.~J. Pearlman, J.~Chen, J.~Fan, A.~Migdall, and J.~P.
  Dowling.
\newblock Resolution and sensitivity of a {Fabry-Perot} interferometer with a
  photon-number-resolving detector.
\newblock {\em Phys. Rev. A}, 80:043822, 2009.

\bibitem{MatekoleEtAlDowling2017-RmTmpPhNrDtctn}
E.~S. Matekole, D.~Vaidyanathan, K.~W. Arai, R.~T. Glasser, H.~Lee, and J.~P.
  Dowling.
\newblock Room-temperature photon-number-resolved detection using a two-mode
  squeezer.
\newblock {\em Phys. Rev. A}, 96:053815, 2017.

\bibitem{HibinoEtAlHofmann2018-StatFromAction}
K.~Hibino, K.~Fujiwara, J.-Y. Wu, M.~Iinuma, and H.~F. Hofmann.
\newblock Derivation of the statistics of quantum measurements from the action
  of unitary dynamics.
\newblock {\em The European Physical Journal Plus}, 133(3):118, 2018.

\bibitem{OkamotoEtAlTakeuchi2008-HsnbgLmt}
R.~Okamoto, H.~F. Hofmann, T.~Nagata, J.~L. O\'Brien, K.~Sasaki, and
  S.~Takeuchi.
\newblock Beating the standard quantum limit: phase super-sensitivity of
  n-photon interferometers.
\newblock {\em New Journal of Physics}, 10(7):073033, 2008.

\bibitem{XiangHofmannPryde2013-MltPhPhsSens}
G.~Y. Xiang, H.~F. Hofmann, and G.~J. Pryde.
\newblock Optimal multi-photon phase sensing with a single interference fringe.
\newblock {\em Scientific Reports}, 3:2684--, 2013.

\bibitem{MarianMarian1993-SqStThmlNoiseI}
P.~Marian and T.~A. Marian.
\newblock Squeezed states with thermal noise. {I. Photon-number statistics}.
\newblock {\em Phys. Rev. A}, 47:4474--4486, 1993.

\bibitem{MarianMarian1993-SqStThmlNoiseII}
P.~Marian and T.~A. Marian.
\newblock Squeezed states with thermal noise. {II. Damping and photon
  counting}.
\newblock {\em Phys. Rev. A}, 47:4487--4495, 1993.

\bibitem{ParisEtAlSiena2003-PurityGaussStInNsyChnnl}
M.~G.~A. Paris, F.~Illuminati, A.~Serafini, and S.~De~Siena.
\newblock Purity of gaussian states: Measurement schemes and time evolution in
  noisy channels.
\newblock {\em Phys. Rev. A}, 68:012314, 2003.

\bibitem{SerafiniEtAlSiena2005-QDecohCVSys}
A.~Serafini, M.~G.~A. Paris, F.~Illuminati, and S.~D. Siena.
\newblock Quantifying decoherence in continuous variable systems.
\newblock {\em Journal of Optics B: Quantum and Semiclassical Optics},
  7(4):R19, 2005.

\bibitem{MonrasParis2007-QEstOfLoss}
A.~Monras and M.~G.~A. Paris.
\newblock Optimal quantum estimation of loss in bosonic channels.
\newblock {\em Phys. Rev. Lett.}, 98:160401, 2007.

\end{thebibliography}
\end{document}